\documentclass[manuscript]{aastex}

\shorttitle{Structure in Dusty Discs}
\shortauthors{Boley et al.}

\usepackage[margin=1in]{geometry} 
\usepackage{amsmath,amsthm,amssymb}
\usepackage{subfigure}
\usepackage{footnote}
\usepackage{ulem}
\usepackage{color}
\usepackage{graphicx}
\usepackage{times} 
\usepackage{amssymb}
\usepackage{amsmath}
\usepackage{lscape}
\usepackage{url}
\usepackage{multirow}
\usepackage{longtable}
\usepackage[english]{babel}
\usepackage{float} 
\usepackage{natbib}

 \graphicspath{{./../files/}}
 \usepackage[table]{xcolor}
 
 \newcommand{\ACBc}[1]{{\color{black}{#1}}}

\begin{document}


\title{On the Origin of  Banded Structure in Dusty Protoplanetary Discs: HL Tau and TW Hya}


\author{A.C. Boley\altaffilmark{1}}
\affil{Department of Physics and Astronomy, University of British Columbia, Vancouver BC, Canada}

\begin{abstract}

Recent observations of HL Tau revealed remarkably detailed structure within the system's circumstellar disc. A range of hypotheses have been proposed to explain the morphology, including, e.g., planet-disc interactions, condensation fronts, and secular gravitational instabilities. While embedded planets seem to be able to explain some of the major structure in the disc through interactions with gas and dust, the substructure, such as low-contrast rings and bands, are not so easily reproduced. Here, we show that dynamical interactions between \ACBc{three planets (only two of which are modelled)} and an initial population of large planetesimals can potentially explain both the major and minor banded features within the system. In this context, the small grains, which are coupled to the gas and reveal the disc morphology, are produced by the collisional evolution of the newly-formed planetesimals, which are ubiquitous in the system and are decoupled from the gas. 

\end{abstract}



\section{Introduction}

The HL Tau system offers a snapshot of the planet formation process.  
This pre-main sequence star is surrounded by a gaseous and dusty disc that has been observed by ALMA at unprecedented sensitivity and resolution \citep{alma_partnership_hltau}.  
The millimetre/submillimetre observations reveal concentric dark and bright rings that appear to be consistent, at least qualitatively, with planets clearing out material, directly reminiscent of planet formation simulations.
Recent calculations and simulations have confirmed that the ring structures can indeed be matched to various degrees by embedded planets \citep[e.g.,][]{dipierro_etal_2015}.   

\ACBc{While the planet hypothesis is compelling and is the focus of this work,  the morphology of HL Tau is not generally accepted to be the result of planets.}
Because the HL Tau system is thought to be $< 1$ Myr old, the existence of planets in the disc  would challenge planet formation theory, as the age of the disc is seemingly too young to have formed planets through core nucleated instability \citep[e.g.,][]{pollack_etal_1996} and the azimuthal symmetry appears to be inconsistent with direct formation in a fragmenting disc \citep[e.g.,][]{durisen_etal_2007,helled_etal_2014}. 
Core nucleated  instability is especially difficult at large stellocentric locations \citep{dodsonrobinson_etal_2009}, such as those exhibited by the structure in HL Tau's disc. 
These reasons, at least in part, have motivated proposals for alternative mechanisms to produce the global morphology.  
This includes, for example, pebble formation at the locations of condensation fronts  of astrophysical ices \citep{zhang_etal_2015} and secular gravitational instabilities \citep{takahashi_etal_2016}. 
\ACBc{Nevertheless, we are strongly motivated by the apparent resonant structure among the rings and gaps in the disc \citep{alma_partnership_hltau} and further argue that the observed morphology has a dynamical, planetary origin.  }

\begin{figure}
\includegraphics[width=0.5\textwidth]{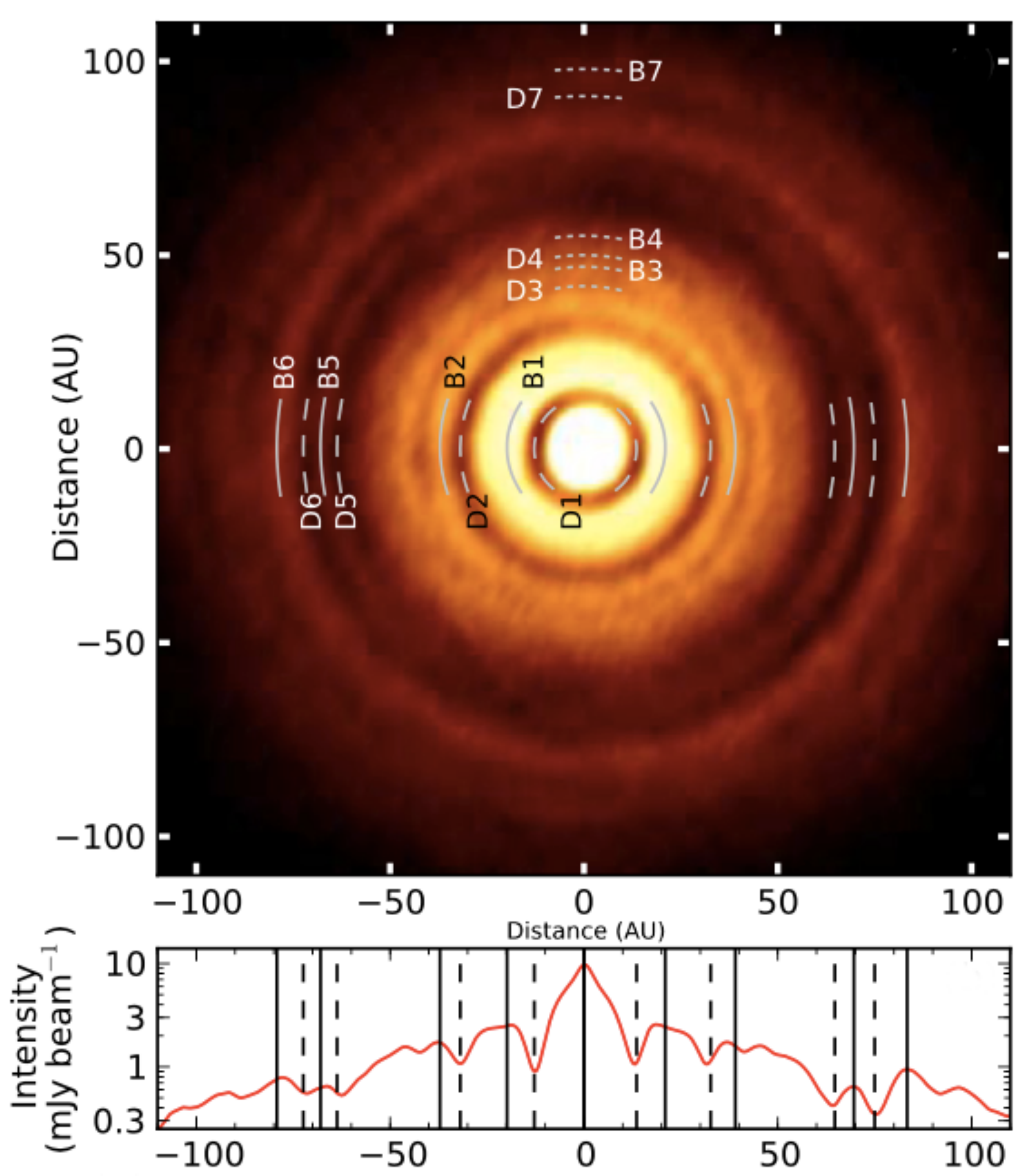}
\caption{\ACBc{Adapted from Figure 3, panels a and b, of the \cite{alma_partnership_hltau}.  {\it Top panel.} ALMA 1 mm (combined bands 6 and 7) continuum image of HL Tau deprojected and annotated with the locations of the bright rings (B) and the dark bands (D) as determined by the ALMA Partnership.    {\it Bottom panel.} Intensity profile of the disc at 1 mm for a slice through the sky projected major axis of the ALMA image (PA=138$^\circ$ in Figure 2 of the ALMA Partnership).  We adopt the ALMA Partnership's naming convention for the rings throughout this paper.  As  discussed in the text, experiments are run with planets in the locations of D2 and B5.  While the paradigm presented here would also suggest that D1 should harbour a planet, we do not include it in our calculations.  The locations of rings and bands that are relevant to this work are listed in Table \ref{tbl:commens}. Please note that asymmetries in the azimuthal appearance of the rings may be due to a loss of image fidelity and should be treated with caution. The solid and dashed lines in the bottom panel show the locations for the image-fitted rings and bands, respectively, as determined by the ALMA Partnership. }
}\label{fig:hltau}
\end{figure}

Figure \ref{fig:hltau} shows the deprojected continuum image with annotations to emphasize the detailed disc structure  \citep[Figure 3a,b from][]{alma_partnership_hltau}.  
Dark bands (simply bands hereafter) are labelled with a ``D'' and are numbered sequentially by stellocentric distance, while the bright rings (rings hereafter) are denoted similarly, but with a ``B''.
As pointed out by the \cite{alma_partnership_hltau}, multiple rings and bands appear to be near commensurabilities. 
While such commensurabilities should be treated with caution \citep{tamayo_hltau}, they potentially provide a wealth of constraints for testing hypotheses on the  disc structure origins.  
For example, let D1 and D2 be the locations of planets.  
Furthermore, let an additional planet be located at B5.  
The motivation for the latter assertion is that D5, B5, and D6, when taken together, appear to be a horseshoe structure,  as is typical of dynamical simulations with planetesimals. For a discussion on this and other plausible arrangements, see \cite{tamayo_hltau}. 
In this proposed configuration, B5:D2 are very close to a 3:1 commensurability.  
D2:D1 are somewhat close to a 4:1, but are notably inside the commensurability.

Moving forward, we envisage that HL Tau does indeed have planets located at D1, D2, and B5.  However, because the innermost regions are poorly resolved, we restrict our discussion to the structure exterior to B1.  This restriction will not change the overall effects, and will simplify the simulations discussed in section 2. For this discussion, we will refer to each planet as PD2 and PB5, respectively.  

Table 1 shows the potential locations for mean motion resonances (MMR) with the hypothesized planets PD2 and PB5.  All band and ring stellocentric distances are taken from the \cite{alma_partnership_hltau}.  
Bands D3 and D4 correspond to a 3:2 and 2:1 with PD2.  The bands D3 and D4 also approximately line up with the 2:1 and 3:2 interior to PB5, while band D7 corresponds to the exterior 3:2 with PB5.

It is well-established that planet-disc interactions can lead to observable features in discs \citep{bryden_etal_1999}, although observing the resulting structure can be non-trivial due, in part, to the range of coupling between gas and dust grains \citep[e.g.,][]{birnstiel_etal_2015}.
The coupling itself depends on the grain sizes as well as the local gas conditions.   
Roughly, small grains remain coupled to the gas, large planetesimals are decoupled, and sizes within the \ACBc{mm to km} range are expected to migrate radially due to gas drag \citep[e.g.,][]{adachi_etal_1976,weidenschilling1977,haghighipour_boss_2003}\ACBc{, although the relevant size range that undergoes significant migration depends on the details of the disc structure.}
As a result, the continuum structure revealed by dust may not represent the actual gas distribution. 
This was highlighted by, e.g., \cite{dipierro_etal_2015}, who ran a series of simulations with different degrees of gas-dust coupling using a range of dust sizes.  
  
Massive planets can open a gap in the gaseous disc whenever waves launched by the planet become non-linear and dissipate angular momentum locally.  
The degree to which gas can be depleted around a planet depends on local disc parameters such as the viscosity and vertical scale height, as well as the planet's mass and the duration of the interactions \citep{kley_nelson_2012}. 
This process will also open a gap in the dust disc for grains $\lesssim \rm mm$, depending on disc location.
For the HL Tau system, a number of studies suggest that such gaps would be opened and potentially consistent with the major bands in HL Tau's disc for planet masses in the range of approximately 0.1 to 1 $M_{\rm jup}$ \citep{dipierro_etal_2015, akiyama_etal_2016, jin_etal_2016, kanagawa_etal_2016}.

While such gap-opening effects would be able to explain D1 and D2, as well as the horseshoe region D5-B5-D6 (depending on the grain size distribution), \ACBc{planets at these locations cannot obviously explain the additional band features such as D3, D4, and D7 through, e.g., resonances if only small grains are present in the gaseous disc.  }
Gas drag effects on millimetre grains are too efficient, and the apparent resonant structure would not manifest.  Therefore, if we consider only the dynamics of gas and millimetre grains, then additional planets must be located at the other bands \citep[e.g.,][]{simbulan_etal_2017}, or a non-planetary origin is required for some of the structure.  On the other hand, we show that the entire detailed structure of HL Tau can be explained by assuming that (1) there are three embedded planets (we focus on only two), that (2) planetesimal formation was, for whatever reason, ubiquitous throughout the disc, and that (3) the millimetre grains observed by ALMA are direct tracers of the dynamically cold planetesimal population. \ACBc{We will  provide physical context for the latter as we present the results.}  We further extend the analysis to TW Hya, in which similar substructure should be seen in more sensitive observations of the disc if  the paradigm presented here is correct. 

\begin{table}
\begin{center}
\caption{The locations of rings and bands in au as determined by  the \cite{alma_partnership_hltau}, along with the locations of commensurabilities with proposed planets.  Bright rings are listed with a ``B'' and dark bands are denoted with a ``D''.  Rows with a ``(P)'' show the locations of proposed planets.  The greater and less than symbols for PB5 are used to highlight commensurabilities interior and exterior to the proposed planet.  The different shades of grey are used to highlight associations with particular commensurabilities.  SMA is the semi-major axis. }\label{tbl:commens}
\begin{tabular}{ccccc}
Band/Ring & SMA & 4:3 & 3:2 & 2:1 \\\hline
(P)D2 & $32.3\pm 0.1$ & $39.1$ & \cellcolor[gray]{0.9}$42.3$ & \cellcolor[gray]{0.7}$51.3$ \\
D3 & \cellcolor[gray]{0.9} $\sim42$ & ... & ... & ...\\
D4 &\cellcolor[gray]{0.7} $\sim50$ & ... & ... & ...\\
D5 & $64.2\pm0.1$ & ... & ... & ... \\
(P)B5$_<$ & \multirow{2}{*}{$68.8\pm 0.1$} & 56.8 & \cellcolor[gray]{0.7}52.5 & \cellcolor[gray]{0.9}43.3 \\
(P)B5$_>$ & & 83.3 & \cellcolor[gray]{0.5}90.1 & 109.2 \\
D6 & $73.7\pm0.1$ & ... & ... & ...\\
D7 & \cellcolor[gray]{0.5}$\sim91$& ... & ... & ...\\
\hline
\end{tabular}
\end{center}
\end{table}

\section{Methodology}\label{sec:method}

We seek to explain the detailed ring/band structure with a flexible, simple processes that could occur in essentially any planet-forming environment.  
We focus the discussion on HL Tau, but will return to TW Hya in section \ref{sec:otherdiscs}.
We further begin by only considering planets on \ACBc{initially} circular orbits, although we will relax this restriction in later sections.  
If a planet is located at B5, as we posit, then its mass needs to be such that a large horseshoe region is produced spanning from D5 to D6.  
\ACBc{Excluding tadpole and horseshoe orbits, as well as gas drag effects,} we know that a planetesimal will become significantly perturbed if it is located a distance $\Delta<2.4 a_p\mu^{1/3}$ from the planet \citep{gladman1993}, where $\mu=M_p/M_*$ (planet to star mass) and $a_p$ is the planet's semi-major axis.  
Throughout these calculations, we assume that the central mass for the HL Tau system is $1.3~M_\odot$. 
We set $\Delta= 8$ au for PB5 at 68.8 au based on the surface brightness profile  \citep{alma_partnership_hltau}, which yields a mass $M_{\rm PB5}=1.5\times10^{-4} M_\odot$ ($\sim 50 M_\oplus$).
For the inner planet at 32.3 au, using the D2 gap profile alone is not a strong constraint on D2's true width due to the narrow angular size of the gap and the size of the beam.  
Instead, we set the inner planet to be approximately a Neptune mass such that $M_{\rm PD2}=5.2\times10^{-5}M_\odot$ ($\sim 17 M_\oplus$). 
This yields a gap size that is reasonably consistent with the width of D2.    
HL Tau does have significant gas present.  
As such, the planet masses should be expected to grow as the system evolves.  
Furthermore, because the planets are already well above $10~M_{\oplus}$, we envisage that growth is currently limited by gas flow into the Hill region rather than the lack of sufficient gas at the location of either planet.

We use {\it Mercury6} with the hybrid integrator to evolve a proposed HL Tau analogue consisting of two planets and 100,000 test particles.  \ACBc{Interactions between the planets are included.  }
The test particles are placed uniformly in semi-major axis between 20 and 100 au, which gives an equivalent surface density profile $\Sigma\propto r^{-1}$, assuming identical particles.
The eccentricities of the test particles are set to zero, but the inclinations are drawn from a uniform random distribution between $0^\circ$ and $1^\circ$.  
\ACBc{The planets are also given  $1^\circ$ inclinations, and the nodes of the planets and test particles are randomized between 0 and $2\pi$.  Including inclinations was done to prevent the simulation from being strictly 2D.}

The orbital time step is set to 140 days, which is used by the hybrid integrator in MVS mode.  The accuracy parameter is set to $1\times10^{-14}$ whenever the hybrid integrator switches to the Bulirsch-Stoer method.  
The simulation is evolved for 100,000 yr.  
\ACBc{While HL Tau could be as old as 1 Myr, the chosen integration time is long enough to allow significant resonant structure in the disc to form.  It further highlights that the mechanism proposed here could, in principle, develop rapidly.}
Each planetesimal is envisaged to be large enough such that gas drag effects are negligible, at least relative to orbital excitation through interactions with planets.  
This is qualitatively consistent with results from planetesimal bow shock studies \citep{morris_etal_2012,hood_weidenschilling_2012}.  

\ACBc{We refer to the above setup as the base HL Tau, base simulation.  For additional simulations, which are used to explore the effects of planet eccentricity and to explore potential signatures in TW Hya, we use the same methodology and summarize the corresponding initial conditions and other changes before discussing the results. }

To justify further the neglect of gas damping for the purposes of these simulations, consider the gas drag force felt by the planetesimal $F_D =\frac{1}{2}C_D\rho \pi R^2 V_{\rm wind}^2$.  Here, the local gas density is $\rho$ and the planetesimal size is $R$.  
The gas-planetesimal relative wind speed is approximated in this case by $V_{\rm wind}\sim e V_{\rm circ}$ for eccentricity $e$ and Keplerian circular speed $V_{\rm circ}$.  
We take the coefficient of drag $C_D\sim 1$ and 
 further let $F_D = m V_{\rm wind}/t_{ D}$ for drag time $t_D$ and planetesimal mass $m$.
 With these assumptions we can write the ratio of the drag time to the local orbital period as 
$\frac{t_D}{T} = \frac{4}{3\pi e C_D } \frac{\rho_p}{\rho}\frac{R}{r}$ at a given orbital distance $r$.  
\ACBc{Setting $R=1$ km, $\rho_p=1\rm~g~cm^{-3}$ (planetesimal internal density), $r=40$ au, $e=0.05$, and $\rho=5\times10^{-13}\rm~g~cm^{-3}$, we find that $\frac{t_D}{T}\sim3000$.   The gas density for the disc is based on the best-fit model from \cite{kwon_etal_2011}.  These values are chosen to emphasize the effects of eccentricity damping on a small planetesimal with modest eccentricity. The distance of 40 au is used to represent roughly the location of B2.  
The resonant forcing timescale is approximately $\sim 100$ orbits, which is described more below. 
Thus, we expect dynamical heating to be effective at the locations of mean motion resonances for planetesimals that are a kilometre in size or larger, while planetesimals that reside away from regions of orbital forcing will remain on cold orbits. }

\ACBc{Radial drift can impact large planetesimals as well as small grains.  Because the drag time is much larger than the local orbital period for kilometric planetesimals, we can write that the change of the planetesimal's orbital angular momentum is $\dot{{\bf L}} = -\frac{1}{2} C_D \rho \pi R^2 V_{\rm wind}^2 r \hat{z}$, assuming a circular, planar orbit.  
In this case, the relative wind speed $V_{\rm wind}$ is the difference between the local gas orbital speed and the planetesimal's orbital speed, which is due to the gas pressure gradient \citep{whipple_1973, adachi_etal_1976, weidenschilling1977}.  
The detailed value of $V_{\rm wind}$ depends on the disc model, but a reasonable estimate is $\sim 5000~\rm cm~s^{-1}$.  
Taking the orbital angular momentum of the planetesimal to be  ${\bf L}=\frac{4\pi}{3}\rho_p R^3 V_c r \hat{z}$ and assuming the size of the planetesimal does not change, the angular momentum can be time differentiated and set equal to the drag torque to find $\dot{r}=-\frac{3}{8\pi} C_D \frac{\rho}{\rho_p} \frac{V_{\rm wind}^2 T}{R}$.  
Using the same conditions as above, the radial drift speed is $\dot{r}\sim 0.2 ~ \rm au~Myr^{-1}$.  
The drift of km-sized planetesimals or larger will thus be overall small.  This is nonetheless dependent on the location and the conditions in the disc.  
During the integration time of the simulations, radial drift could cause some of the km-sized planetesimals to drift into a resonance, although this effect would not be captured by the given simulations. 
On the other hand, planetesimals with larger eccentricities, such as those in mean motion resonances, will experience much larger relative wind speeds and could drift out of the resonant locations.}

\ACBc{ The behaviour of small grains will be different.  The drag time for these particles is given by $t_D\sim \rho_g s /(\rho v_{th})$, where $\rho_g$ is the grain's density, $s$ is the grain radius, $\rho$ is the gas density, and $v_{\rm th}$ is the mean thermal speed of the gas.  If $s\sim 1$ mm, $\rho_g\sim 1\rm~g~cm^{-3}$, and the gas temperature is $\sim 55$K, then again for the conditions at $r=40$ au, $t_D\sim 0.09$ yr.  
As such, these small grains will not respond to resonant eccentricity forcing.
If the grains are suddenly released from an eccentric planetesimal, then the small grains will quickly acquire the local orbital conditions.
Moreover, the difference between the gas's orbital motion and the circular, Keplerian speed will result in a residual radial gravity term $\Delta g$ felt by the grain \citep{weidenschilling1977}.
At 40 au $\Delta g \approx 9\times10^{-6}~\rm cm~s^{-2}$ for the assumed wind speed $V_{\rm wind}=5000~\rm cm~s^{-1}$.  
The resulting radial drift of the mm grains is $\dot{r}\sim \Delta g t_D\sim 50~\rm au~Myr^{-1}$.  
As such, while mm grains will have their eccentricities quickly damped, they will undergo substantial radial migration unless they are reaccreted onto large planetesimals or are destroyed. }

\section{Results}

\subsection{HL Tau, Base Simulation}\label{sec:basesim}

\begin{figure}
\includegraphics[width=0.45\textwidth]{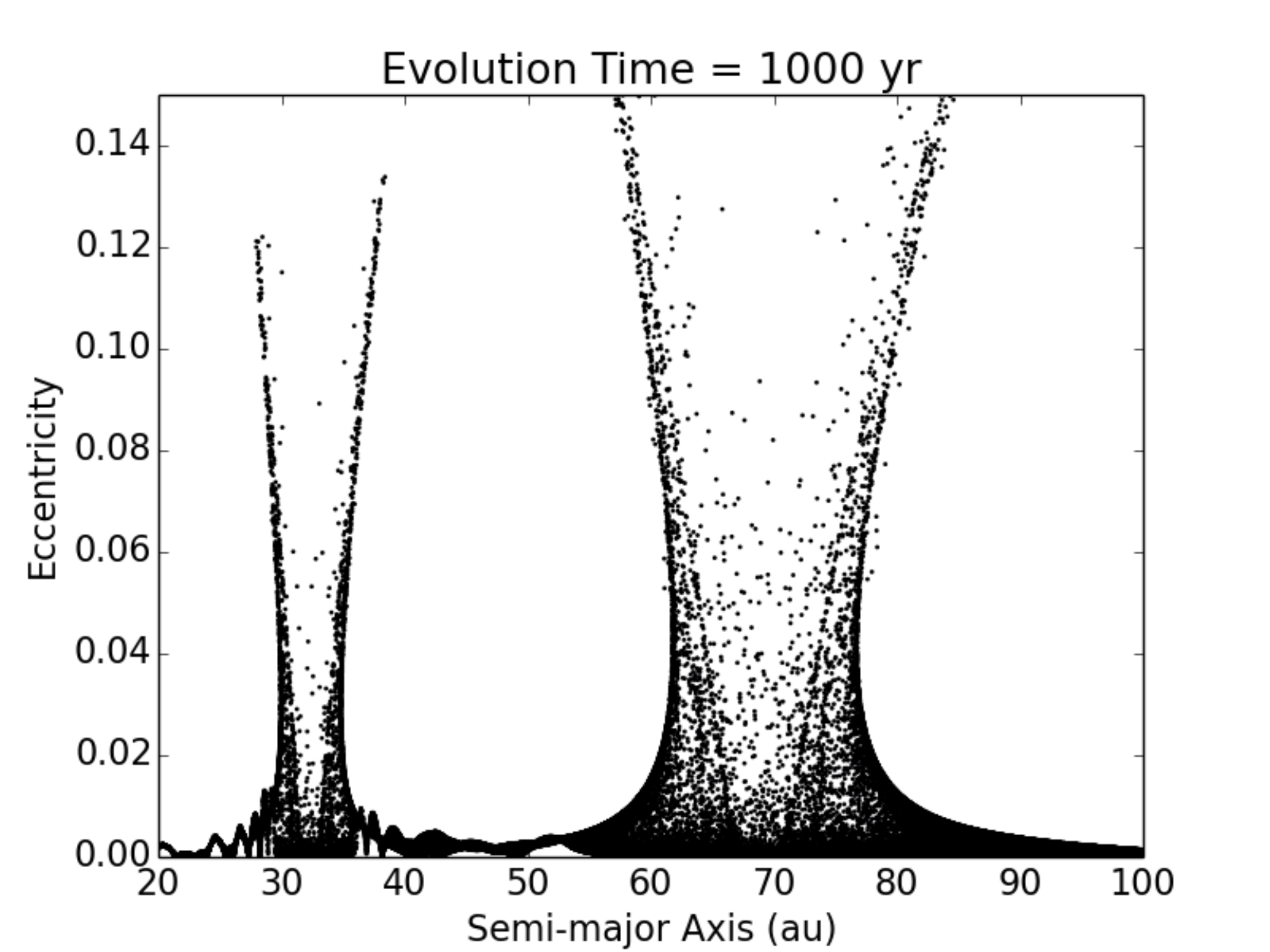}\includegraphics[width=0.45\textwidth]{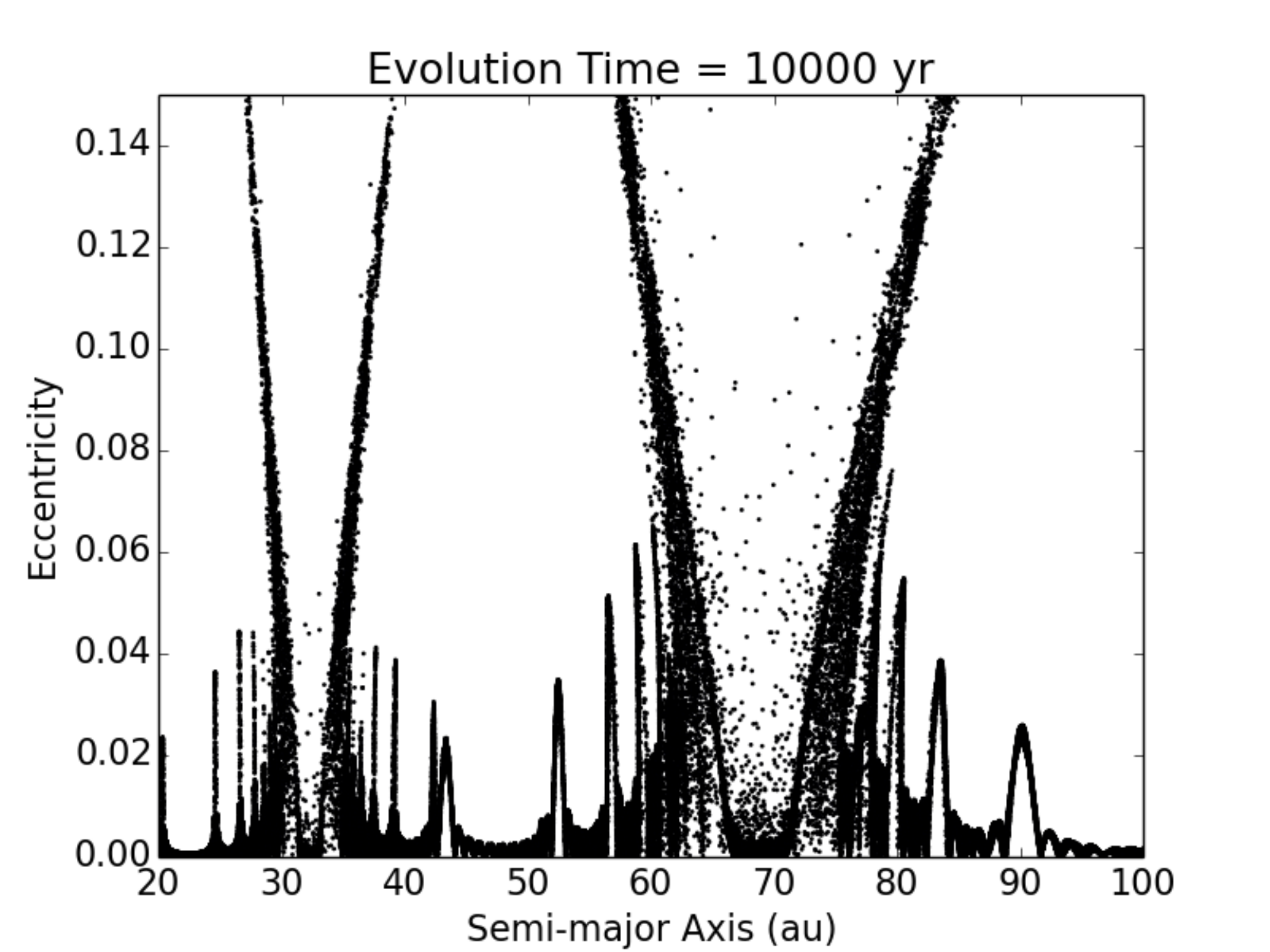}
\includegraphics[width=0.45\textwidth]{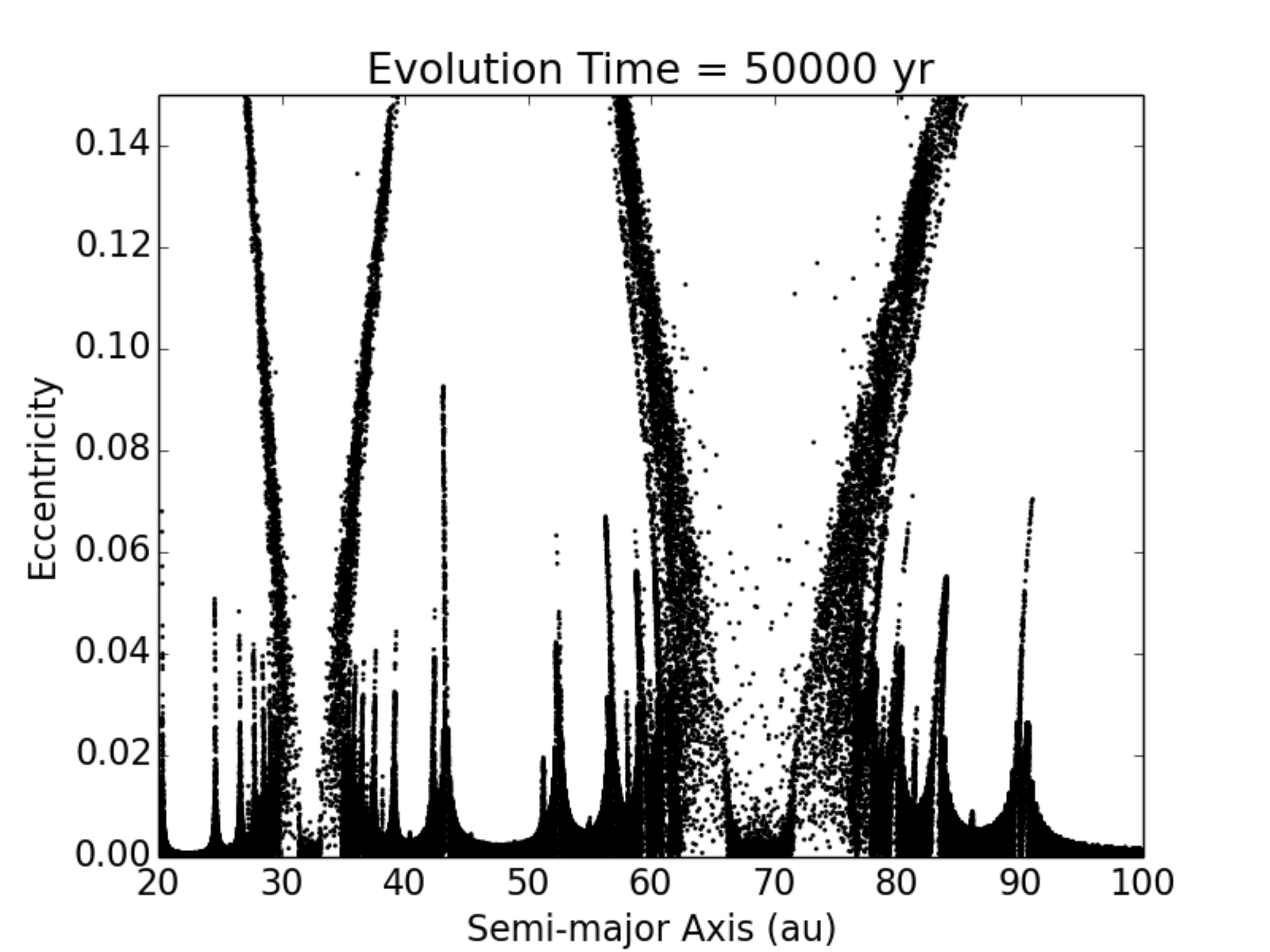}\includegraphics[width=0.45\textwidth]{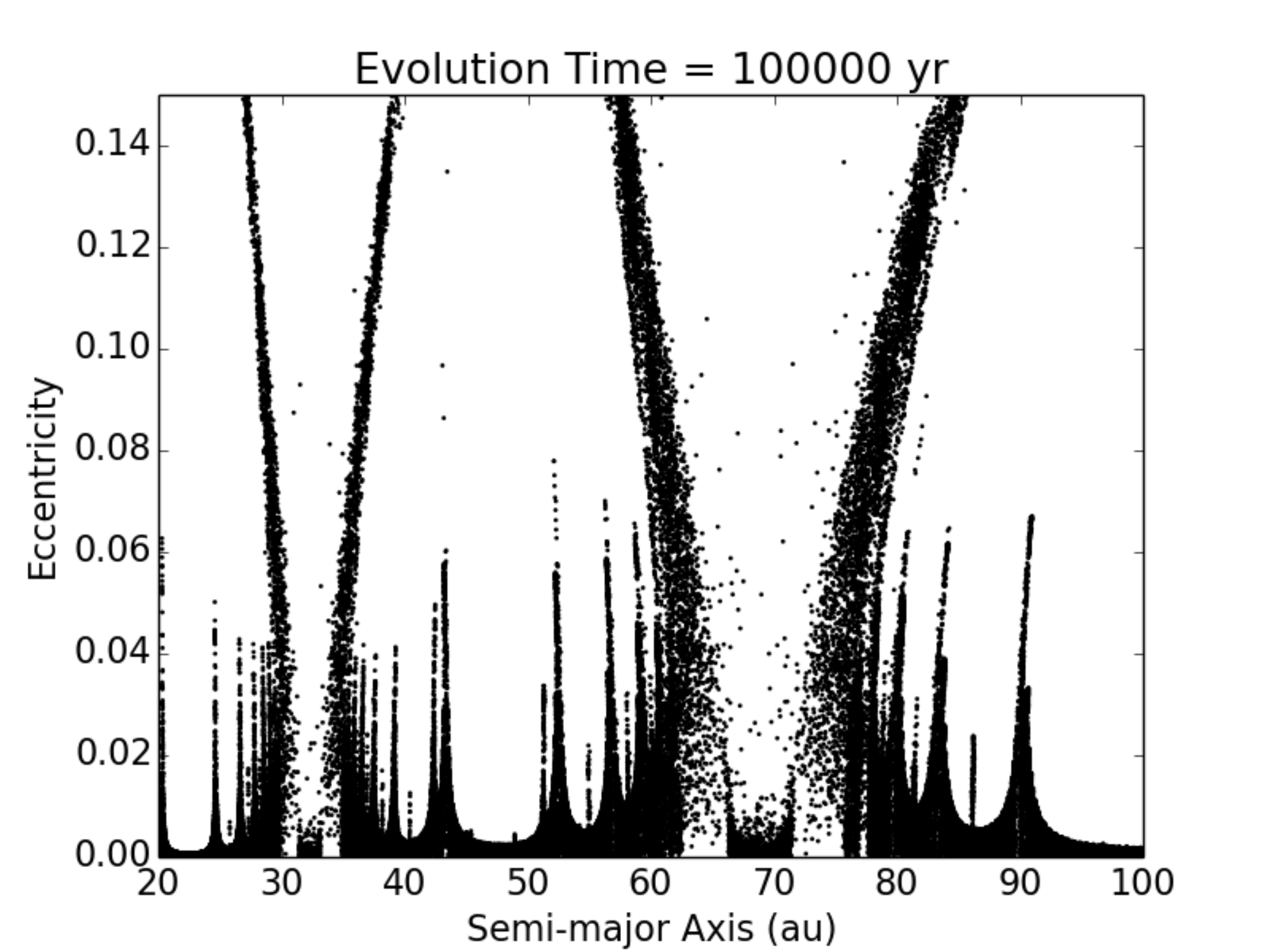}
\caption{Eccentricity evolution of the planetesimals for the HL Tau, base simulation.  The resonant structure forms quickly, consistent with the timescales presented in section \ref{sec:method}.  The figure also shows that $10^5$ yr of evolution  is sufficient to capture the overall resonant structure. 
The high-eccentricity wings near the locations of planets extend to an eccentricity of about 0.3.  
}\label{fig:aetime}
\end{figure}

\ACBc{ The eccentricity evolution of the planetesimals in the HL Tau, base simulation is shown in Figure \ref{fig:aetime}.  After a few 10,000 yr, the main resonant structure is established.   We do not expect the morphology to undergo significant changes for integration times longer than 100,000 yr, at least for the current n-body model. Moving forward, we will take 100,000 yr to be a representative snapshot for the proposed system. }

\begin{figure}
\includegraphics[width=0.45\textwidth]{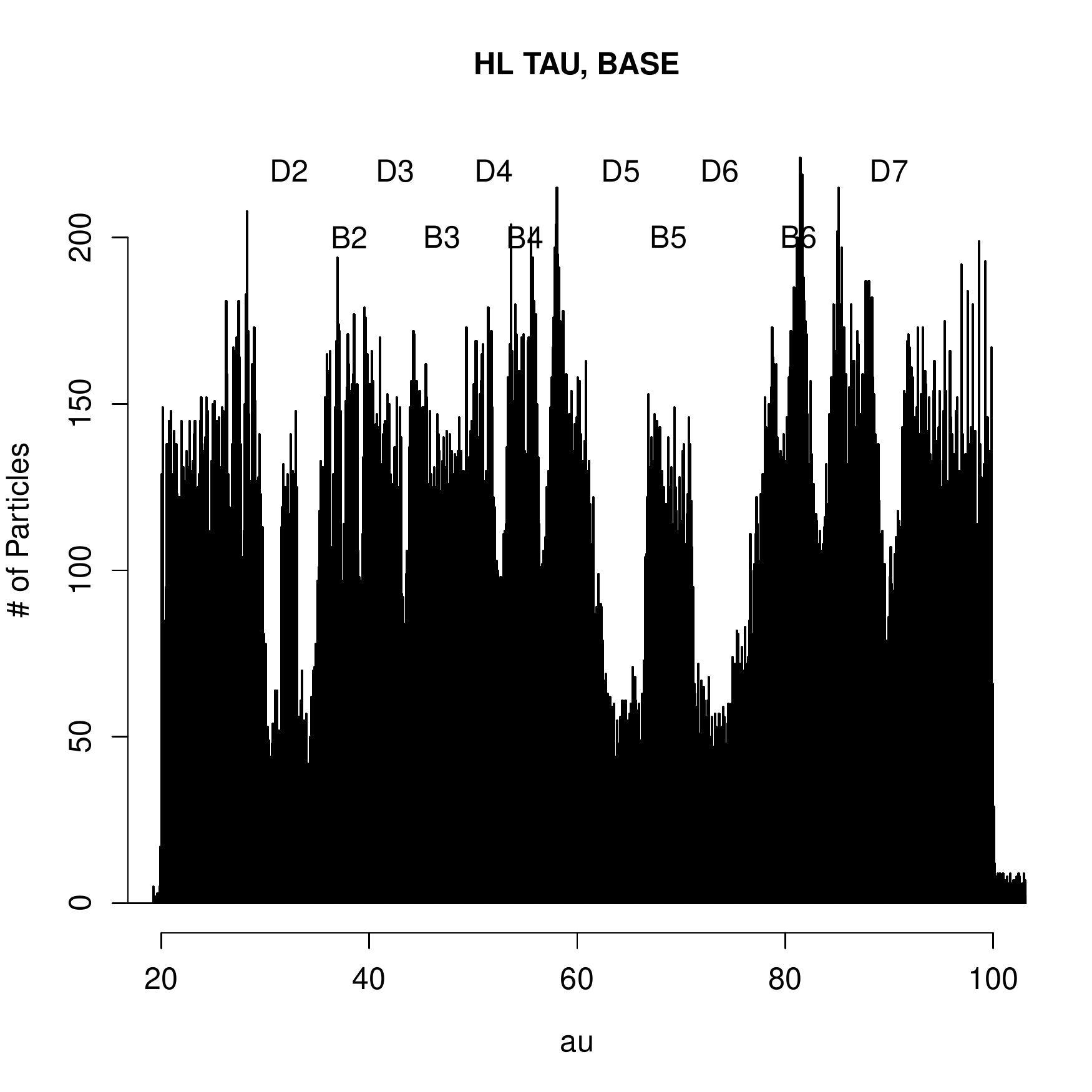}\includegraphics[width=0.45\textwidth]{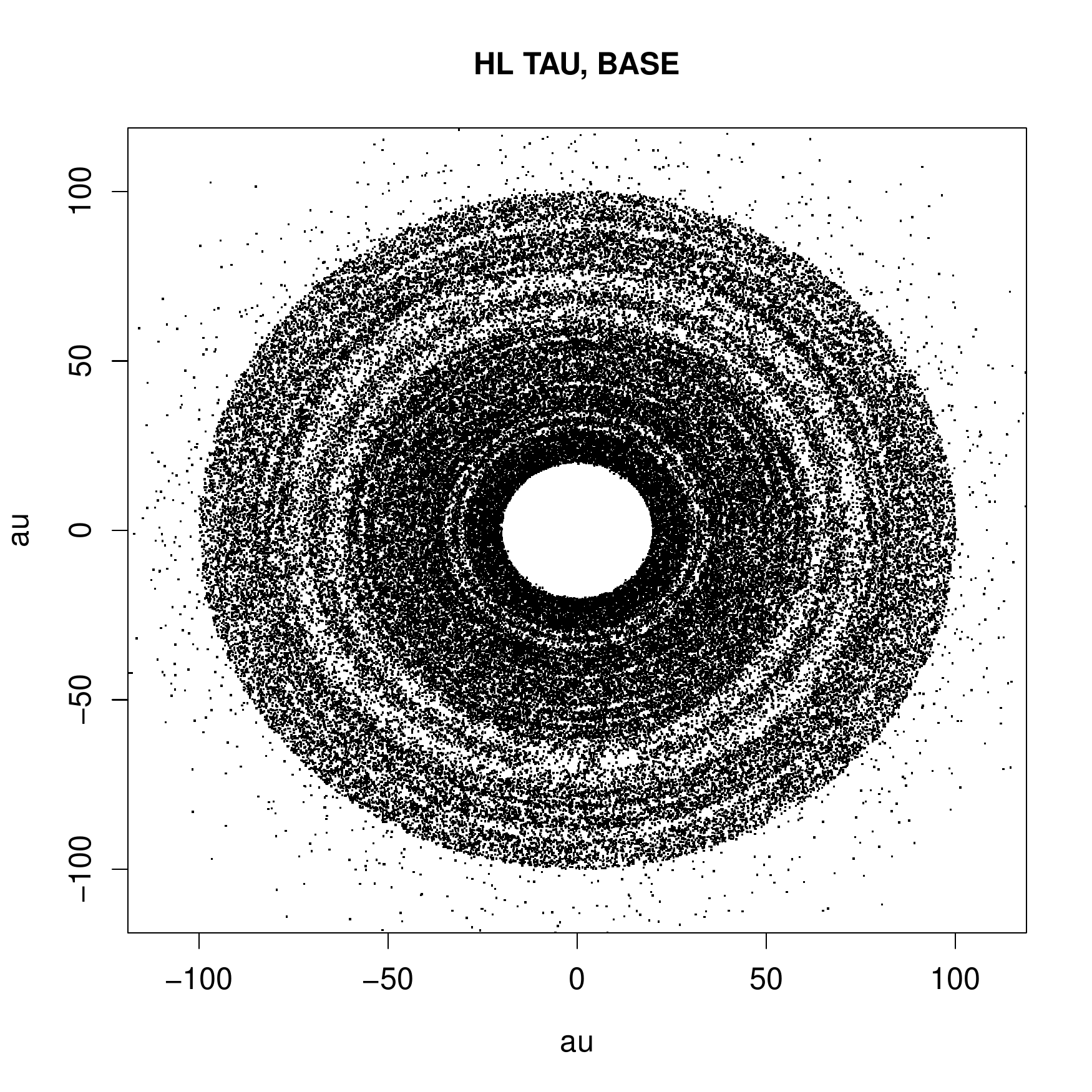}
\caption{Radial binning of the location of the test particles at the end of the base simulation.  Bin widths are set to 0.1 au. The locations of the planets show clear horseshoe gaps and  there is a depletion of test particles at low-order commensurabilities with the planets. Because the planets are near a 3:1 commensurability, the 3:2/1:2 and 2:1/2:3  (planet:planetesimal) commensurabilities associated with the inner and outer planets nearly overlap  (Table \ref{tbl:commens}).  
The labels show the nominal locations of the major rings and bands as determined by the \cite{alma_partnership_hltau}.  
If the given ring or band is only approximately known, the labels have minor shifts to line up with the nearest feature in the simulation.  Despite the clear alignment, the gaps produced at the commensurabilities are neither deep nor are they wide. The labels are centred on radial positions of 32.36  au (D2),  38.1 au (B2), 42.5 au (D3), 47 au (B3), 52 au (D4), 55 au (B4), 64.2 au (D5), 68.8 au (B5), 73.7 au (D6),  81.3 au (B6), 90 au (D7). }\label{fig:hltau_hist}
\end{figure}

Figure \ref{fig:hltau_hist} (left) shows the radial locations of planetesimals after 100,000 yr of evolution.  
Binning is done according to the physical location of the planetesimal, not its semi-major axis.  
Horseshoe gap structures are clearly produced at the location of each planet. 
 In addition, mean motion resonances between the planetesimals and the planets deplete the number of planetesimals that are located near bands in HL Tau, assuming the mm grains that are observed by ALMA follow the planetesimals.  
However, the agreement is insufficient. The gaps are too small to be reliably imaged, and the structure as a whole is not a strong match to the observations.  
This is emphasized in the right panel of the figure, which shows the face-on $X$-$Y$ positions of the planetesimals.  The gaps are present, but are small and are azimuthally asymmetric due to variations in the orbital orientations of the high-eccentricity planetesimals.  Even the horseshoe regions are not so clearly delineated in the face-on image.

We now proceed under the following framework: the small grains observed by ALMA are debris, produced by the collisions among the already-formed planetesimals.  If this is the case, we might expect that the hot population of planetesimals  is unlikely to collide with itself, and if a hot planetesimal does collide with an object, it does so with the cold population due to the higher density \citep[e.g.,][]{wyatt_etal_2010,lawler_etal_2015}.  
The small grain radial drift time through the cold population of planetesimals is assumed to be long compared with the reaccretion timescale of those grains onto different planetesimals due to, e.g., pebble accretion \citep{johansen_etal_2015}.
Under these assumptions \ACBc{(discussed in more detail below)}, small grains will be more strongly associated with the dynamically cold planetesimals, which will serve as a spatial filter.  Throughout the analysis here, we use an eccentricity cutoff of 0.005 for the hot and cold populations.  \ACBc{Except in regions close to the planets, planetesimals with eccentricities larger than this value are associated with mean motion resonances}.

\begin{figure}
\includegraphics[width=0.45\textwidth]{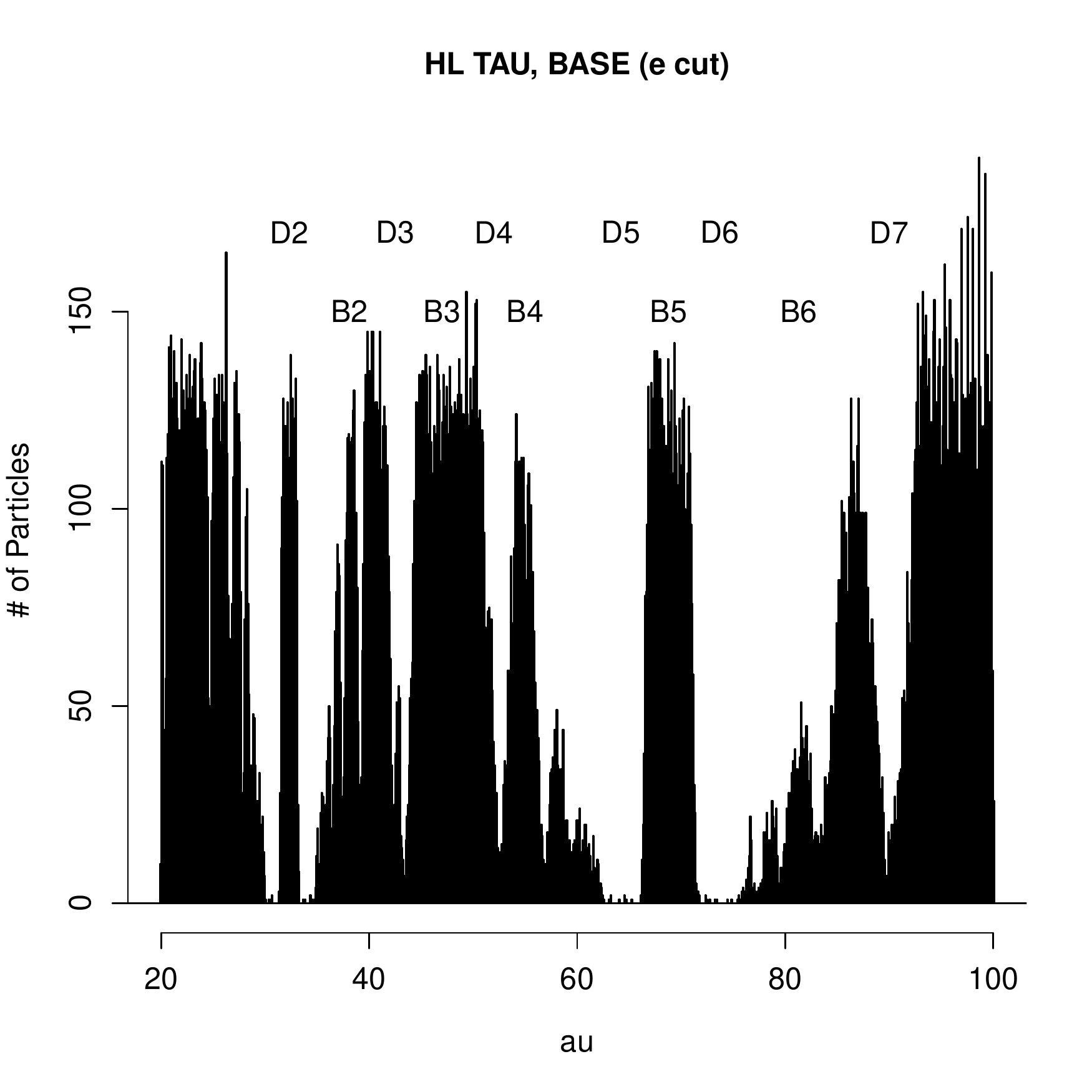}\includegraphics[width=0.45\textwidth]{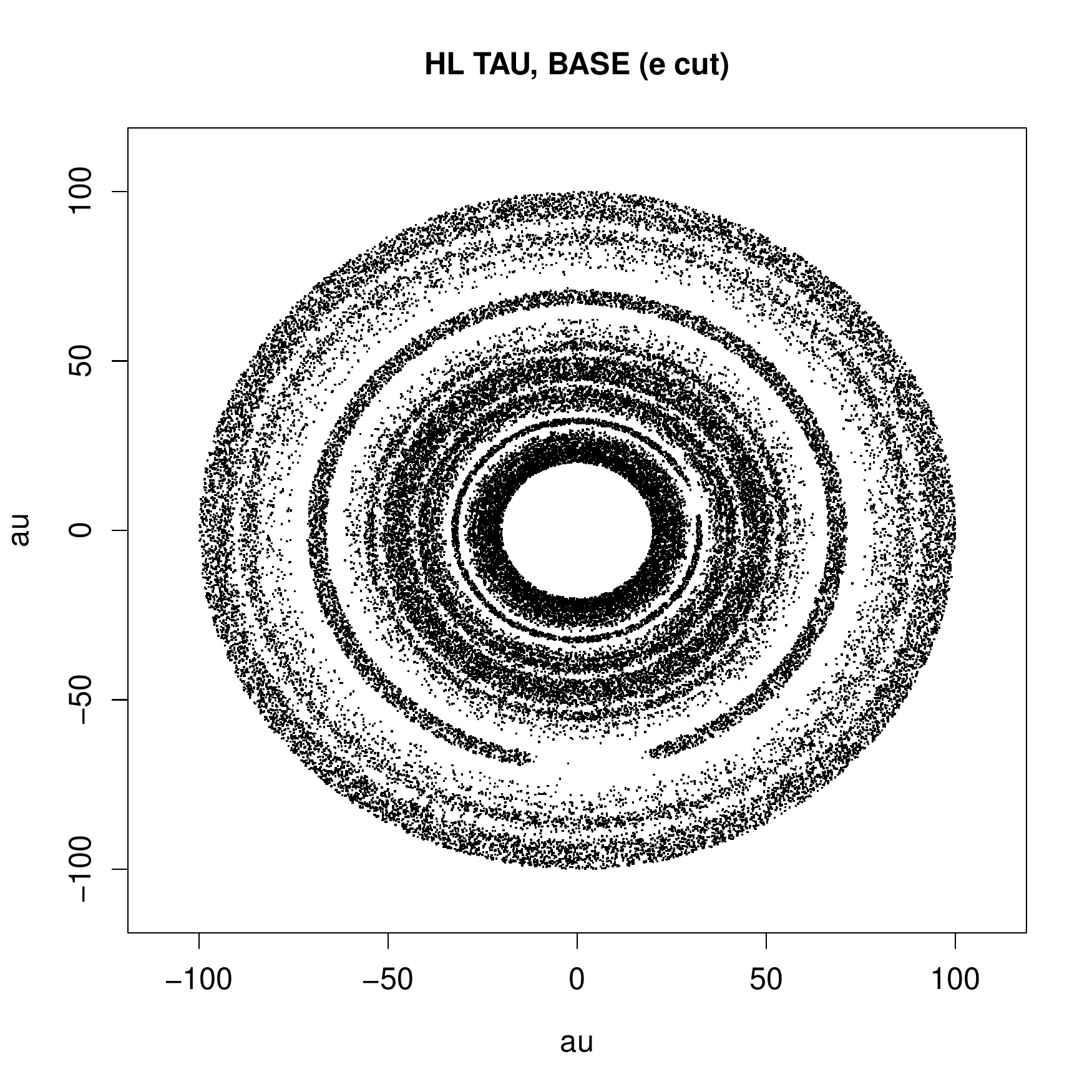}
\caption{Same as Figure \ref{fig:hltau_hist}, but using an eccentricity cutoff of $e=0.005$, for which planetesimals with eccentricities in excess of this limit are excluded in the histogram (left panel) and the face-on positions (right panel). The locations between the rings and bands are consistent with the actual morphology in Figure \ref{fig:hltau}. }\label{fig:hltau_hist_ecc}
\end{figure}

The consequences of our assumptions are highlighted in Figure \ref{fig:hltau_hist_ecc}.
The ring and bands exhibit strong overlap with the  morphological features in the ALMA image, although the observability of the simulated disc will ultimately depend on the detailed collisional model and subsequent dust evolution (under the debris hypothesis).  We note that D2 displays an additional horseshoe region.  If the planets are on circular orbits, as envisaged here, this type of structure must be a result.  Because this is not seen in the current ALMA continuum images, we offer two potential explanations.  (1) The horseshoe gap is present for D2, but the current image is still too low resolution/sensitivity to show this structure.  If this is the case, should more sensitive and higher resolution images become available, we can test for the presence of a horseshoe gap similar to that potentially seen with the D5-B5-D6 morphology.  (2) The planet PD2 has a lower mass than what is used in the base simulation (Neptune-mass) and it is on an eccentric orbit.  We show one such possible configuration in the next subsection.  

There are two additional issues.  First, the location of the B6 ring is potentially at too large of a radius.  The face-on image shows that planetesimals populate the necessary region, but the peak of the distribution is offset due, in part, to the use of an eccentricity cut.  We interpret this to suggest that grain physics cannot be ignored completely and that a detailed collisional model is necessary to determine the exact locations of features.  

The second issue is that the \cite{alma_partnership_hltau} found that the rings in the deprojected ALMA image appear to have a low ($e\lesssim0.033$), but non-negligible eccentricity (based on ring offsets from the 1 mm peak emission).  
If this is not an image artefact, then secondary interactions between the gas and the small grains \citep[e.g., see][]{jin_etal_2016} would be needed to produce such structure, as an eccentric planet mainly makes a planetesimal gap wider rather than eccentric in the proposed context.

Even with these outstanding issues, which ultimately need to be addressed, the structure as a whole is interpreted to have high fidelity with the morphology of the ALMA image.

\subsection{HL Tau, eccentric case}

A planet on an eccentric orbit has several consequences under the proposed paradigm.  First, the gap width would be controlled by the mass of the planet {\it and} by the planet's orbital eccentricity \citep[e.g.,][]{deck_etal_2013}.   We illustrate this case by placing PD2 at a semi-major axis of 33.076 au, which is at the 3:1 commensurability with PB5.  The mass of PD2 is reduced to 1 $M_{\oplus}$, and the orbital eccentricity is increased from circular  to $e=0.005$.  The semi-major axis and mass of PB5 remain unchanged, but its eccentricity is increased to $e=0.001$.  The simulation setup is otherwise unchanged from the circular initial conditions.

\begin{figure}
\includegraphics[width=0.45\textwidth]{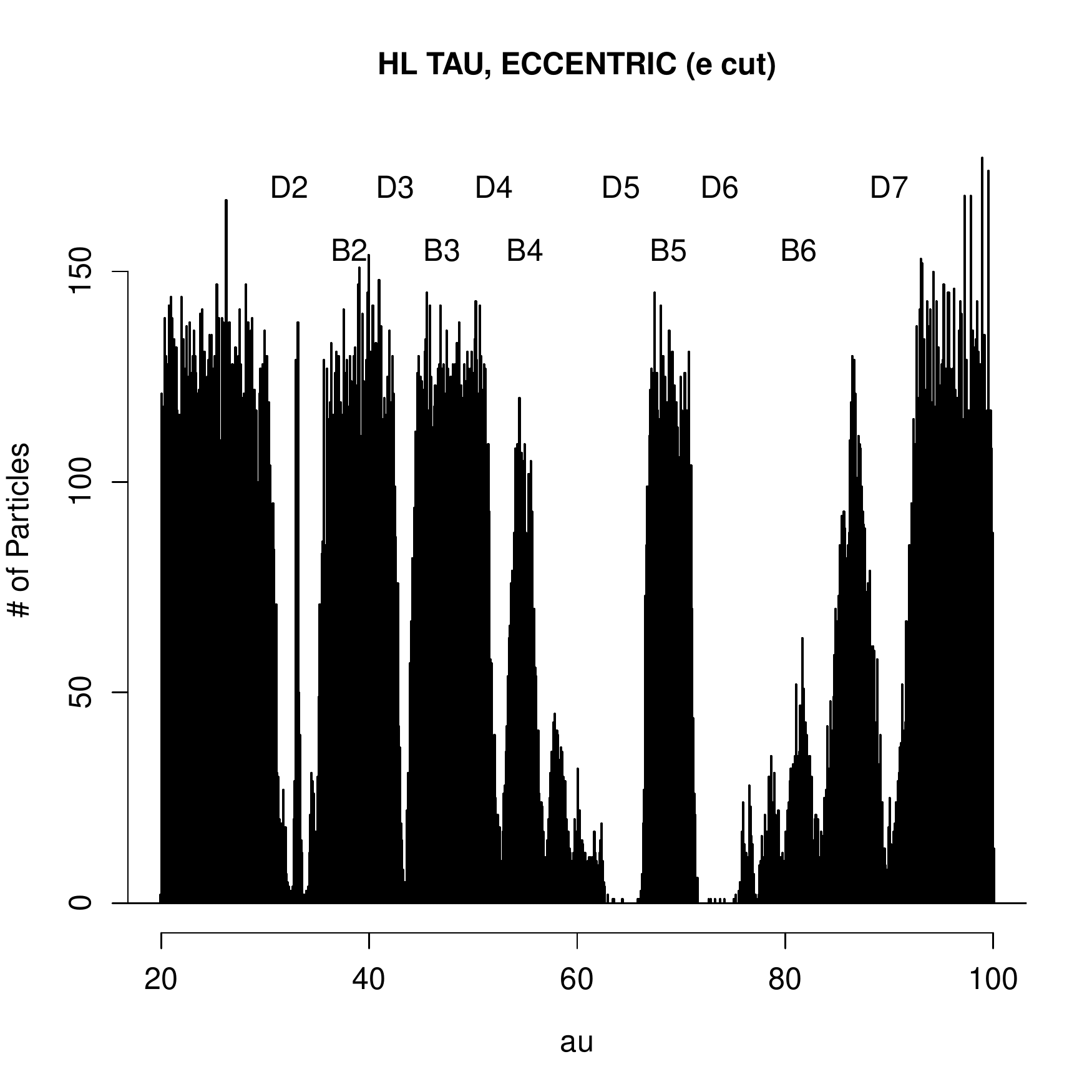}\includegraphics[width=0.45\textwidth]{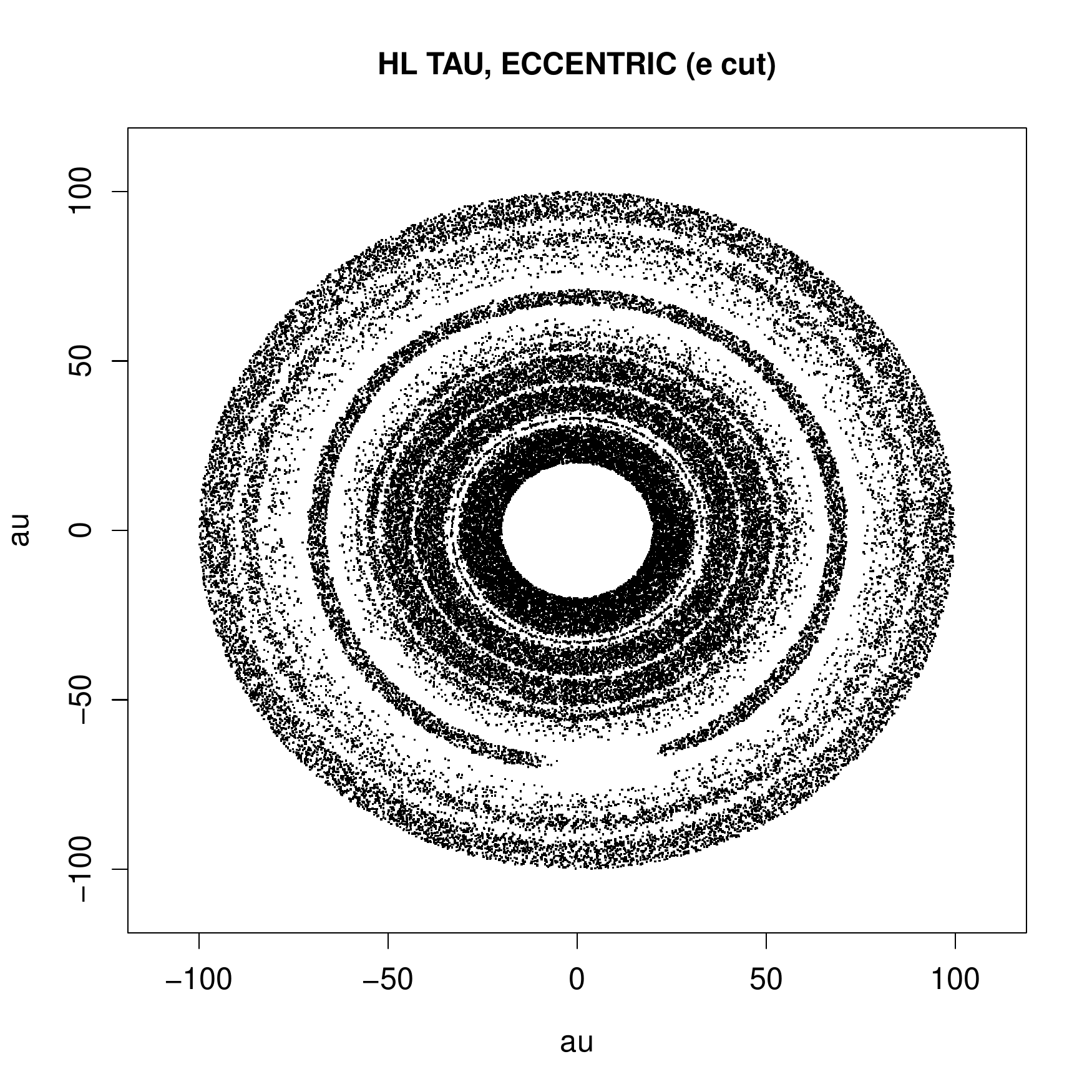}
\caption{Same as Figure \ref{fig:hltau_hist_ecc}, but assuming PD2 is lower mass ($1 M_{\rm \oplus}$) and has an $e=0.005$.  The mass of PB5 is kept the same, but the initial eccentricity is increased to $e=0.001$. The higher eccentricity for PD2 reduces the presence of the horseshoe gap, while also allowing a wide gap with a lower-mass planet.}\label{fig:hltau_v12}
\end{figure}

Without any eccentricity, we would expect the resulting D2 to exhibit a width of approximately 2 au and to have a well-defined horseshoe structure.  With the given eccentricity, PD2 opens D2 to a width of about 3-4 au.  The horseshoe structure, while still present, is greatly reduced and could be missed entirely by observations.  The full structure with the eccentricity cut is shown in Figure \ref{fig:hltau_v12}.

The circular and eccentric cases highlight the variety of conditions that could produce structure within discs, including morphologies with and without horseshoe gaps.   
It does not, however, give rise to limitless possibilities.  Substructure must still be reconciled with the locations of resonances.  Moreover, to be observable, the resonant gaps must have sufficient width.  This may only be possible upon reaching a certain mass range for the planets.

\section{Discussion}

The agreement between the simulation results and the locations of the rings and bands suggest that the structure is consistent with planets located at D2 and B5.  
The model used here posits that the mm grains are created by planetesimals through collisions, as small grains alone would be too coupled to the gas to be affected by the dynamics of the planets, at least in a way that is consistent with the  full ring and band structure. 
Once produced in the disc, the small grains trace the dynamically cold planetesimal population under the assumption that reaccretion of pebbles is faster than radial drift. 
Such a model has multiple consequences.
\begin{itemize}
\item Planetesimal formation is rapid, and occurs throughout the entire disc.  If correct, planetesimals do not, in general,  form at special locations in the disc such as at dead zone boundaries.  We do not address the formation mechanism here, but do argue that the mechanism must be rapid and global in extent. 

\item Small grains, including mm sizes, do not in general reveal grain growth. HL Tau is believed to be less than 1 Myr old.  If the morphology of the disc is mainly due to planetesimal dynamics, rather than small grain dynamics, then the system may be best thought of as a newly-formed debris disc, which is heavily embedded in gas.  This should be a common pathway of evolution for planet-forming discs. 

\item Grains are continuously recycled among the planetesimals.  Collisions among planetesimals will produce grains, which then go through a pebble accretion-like mass redistribution onto the remaining planetesimals.  In this way, large-scale migration of mm-grains is avoided, as the grains spend most of their time locked up in large planetesimals \citep{boley_ford_2013}.  It also keeps the grains largely confined to the location of the parent bodies.  Grains that do form in regions of low planetesimal density may, however, experience migration into areas of higher planetesimal density, where they would be more likely to be accreted. 

\end{itemize}

This mechanism should be general.  The telltale signatures of this process include the following: (1) Horseshoe gaps should be present at the location of a planet, provided that the planet's orbital eccentricity is low.  High resolution interferometry with high image fidelity may be necessary to fully see the horseshoe structure (e.g., D2 vs. D5-B5-D6 in HL Tau).  (2) Discs should exhibit a fine ring and band structure in which minor bands are in commensurabilities with the centres of very deep gaps or with the centre of possible horseshoe structures.  These features should be most noticeable when interior and exterior commensurabilities between the planets slightly overlap, widening the resulting band.  (3) Unless large-scale asymmetries are present due to, e.g., gravitational instabilities \citep[][]{tobin_etal_2016}, all discs should show the ring-band structure while significant gas is present.  Nonetheless, we expect the banded structure to fade with time.  This is due, in part, to planetesimal self-stirring as the gas dissipates, which will cause the cold population to heat dynamically and wash out small radial variations.

\subsection{\ACBc{Assumptions regarding planetesimals and small grains}}

One of the main assumptions used in this work is that the small grains trace the dynamically cold planetesimal population.  
This is envisaged to be due to (1) planetesimal collisions being more likely within the cold population, along with (2) recycling of those grains through reaccretion.  
We explore the collisional assumption in two ways, which  yield comparable results.  

First, let  the collision rate for one planetesimal to hit another be approximated by  \citep[e.g.,][]{safronov_1972}:
\begin{equation}
\dot{N}=\frac{3 \pi R_c^2 \Sigma_d F_g}{2 \rho_p R^3 T},
\end{equation}
for collision radius $R_c$, planetesimal density $\rho_p$, planetesimal size $R$, and orbital period $T$.  
This rate assumes that each planetesimal effectively encounters the local disc surface density of solids $\Sigma_d$  twice per orbit.
The gravitational focusing parameter $F_g$ depends on the properties of the planetesimals and the speed $V_{\rm rel}$ of a given planetesimal relative to the local population.  We assume identical planetesimals, which sets $R_c=2R$;
\begin{equation}\label{eqn:simple_collision}
\dot{N}=\frac{6 \pi \Sigma_d F_g}{ \rho_p R T},
\end{equation}
and
\begin{equation}
F_G = 1 + \frac{8 \pi G \rho_p R^2}{3 V_{\rm rel}^2}.
\end{equation}
We take the final state of  the HL Tau, base simulation, and bin the particles by their semi-major axes.  
For our analysis, we use 0.1 au bin widths.  
Using smaller bins increases the noise in the calculations, while larger bin widths washes out structure.
In each radial bin we calculate a planetesimal velocity dispersion.  
This dispersion is determined by first finding the speed of each particle relative to a circular, planar orbit at the instantaneous cylindrical  distance from the star. 
These relative speeds for each particle are then used in the calculation of the velocity dispersion for the given particle's semi-major axis bin.  
We exclude particles with an eccentricity $e>0.1$.  
Particles with eccentricities larger than this value are almost exclusively associated with the high-eccentricity wings of the planet clearing zones (Figure \ref{fig:aetime}).  
Particles in those wings, which extend over a large range of semi-major axes, skew the averaging used in the dispersion calculation.  

Based on equation (\ref{eqn:simple_collision}), we expect the following behaviour, assuming identical planetesimals.  
When $F_g\sim 1$, the collision rate is only affected by $\Sigma_d/T$.
The collision rates for a sea of small planetesimals will only show strong, local variations if there are corresponding changes in the surface density.    
As gravitational focusing becomes important, then collision rates will increase at semi-major axes that have dynamically cold planetesimals, while rates will remain low in regions with high orbital excitation.   This will give rise to strong local variations that depend  on $F_g$.  
In this way, if the picture presented here is correct, the existence of rings and bands could be a tool for evaluating the size scales of planetesimals in a disc. 

\begin{figure}
\includegraphics[width=0.45\textwidth]{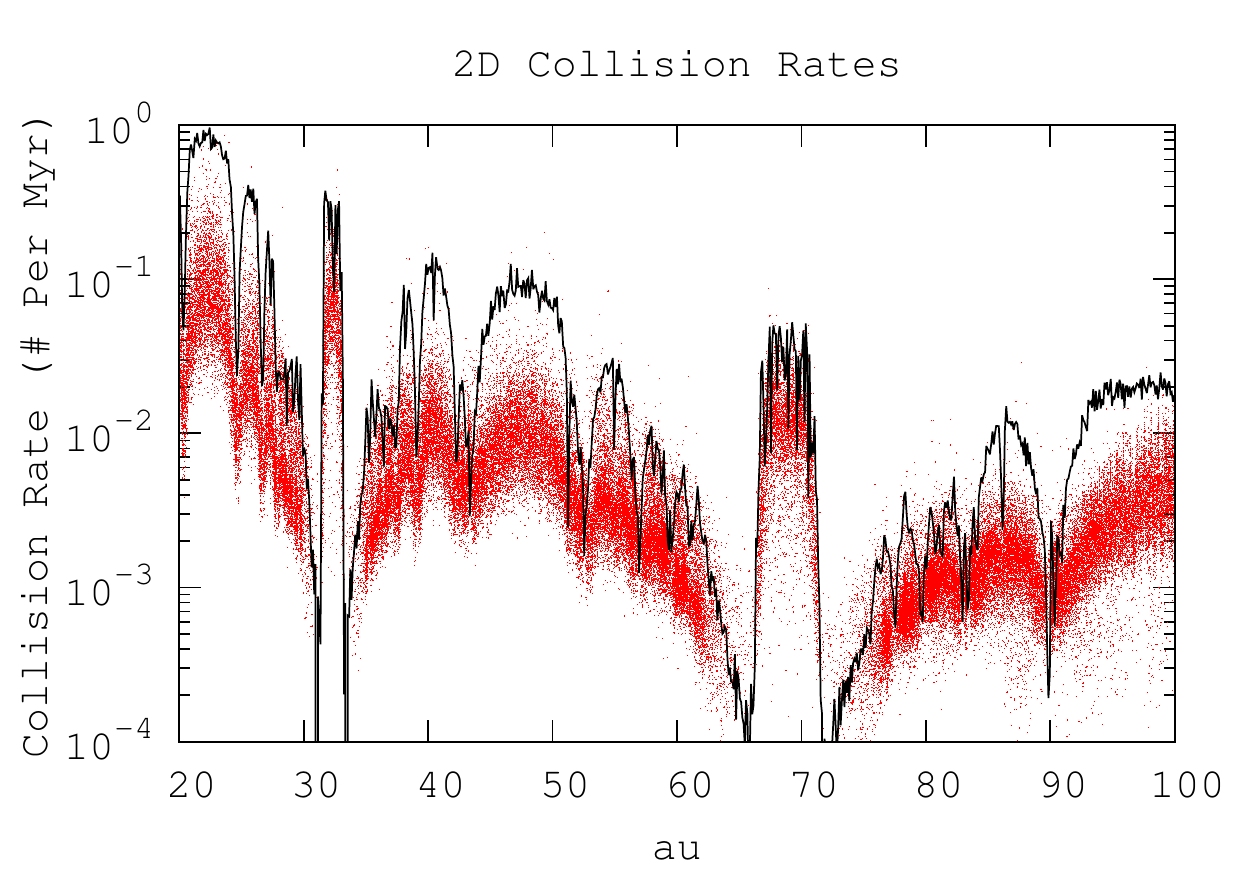}\includegraphics[width=0.45\textwidth]{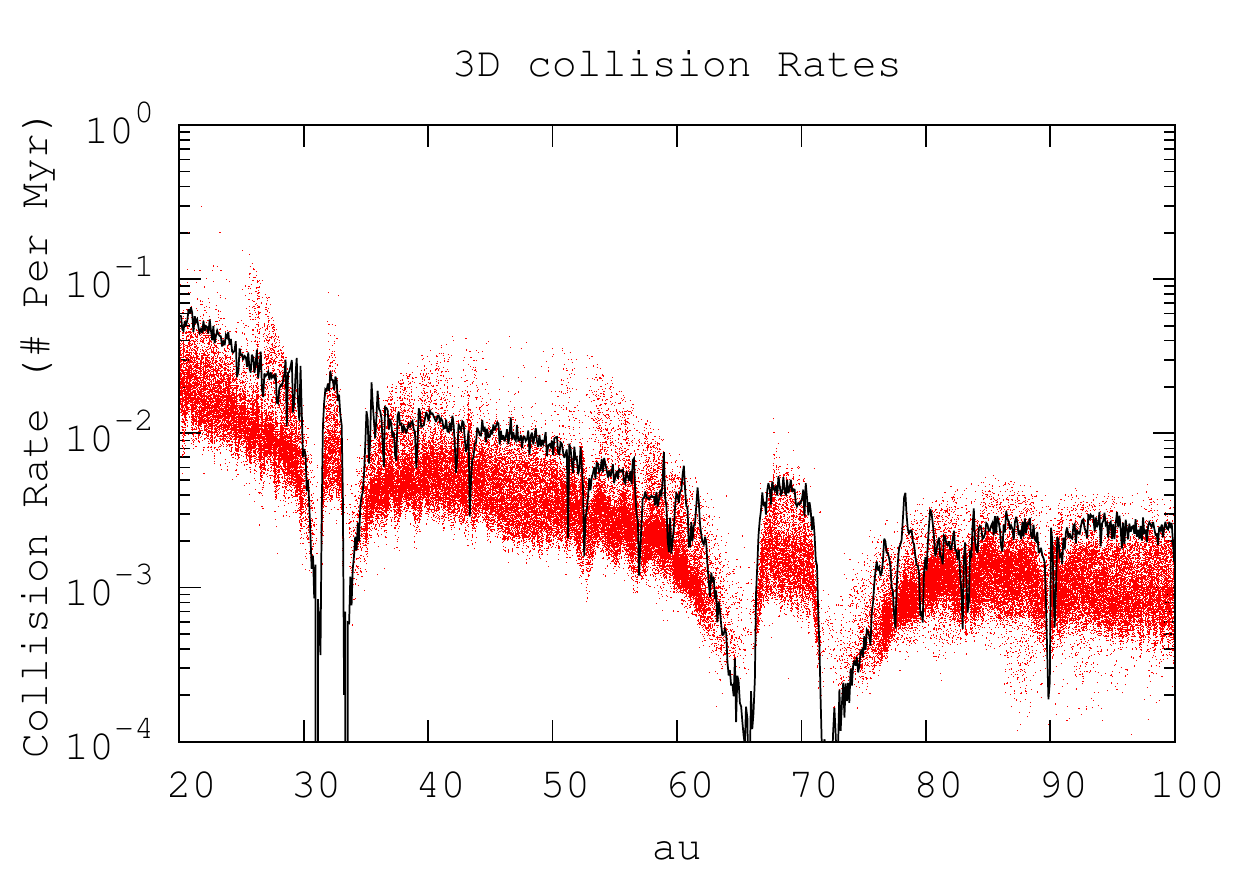}
\includegraphics[width=0.45\textwidth]{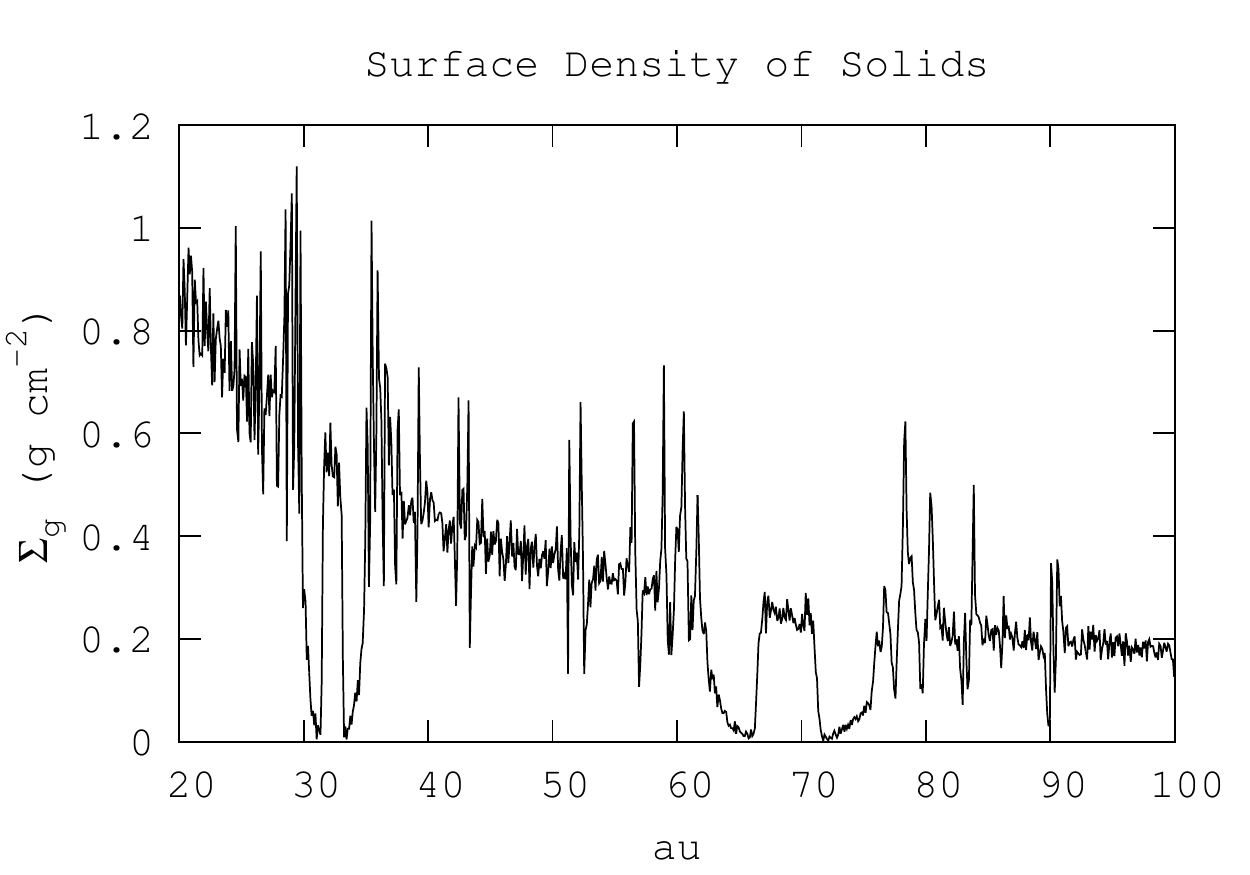}\includegraphics[width=0.45\textwidth]{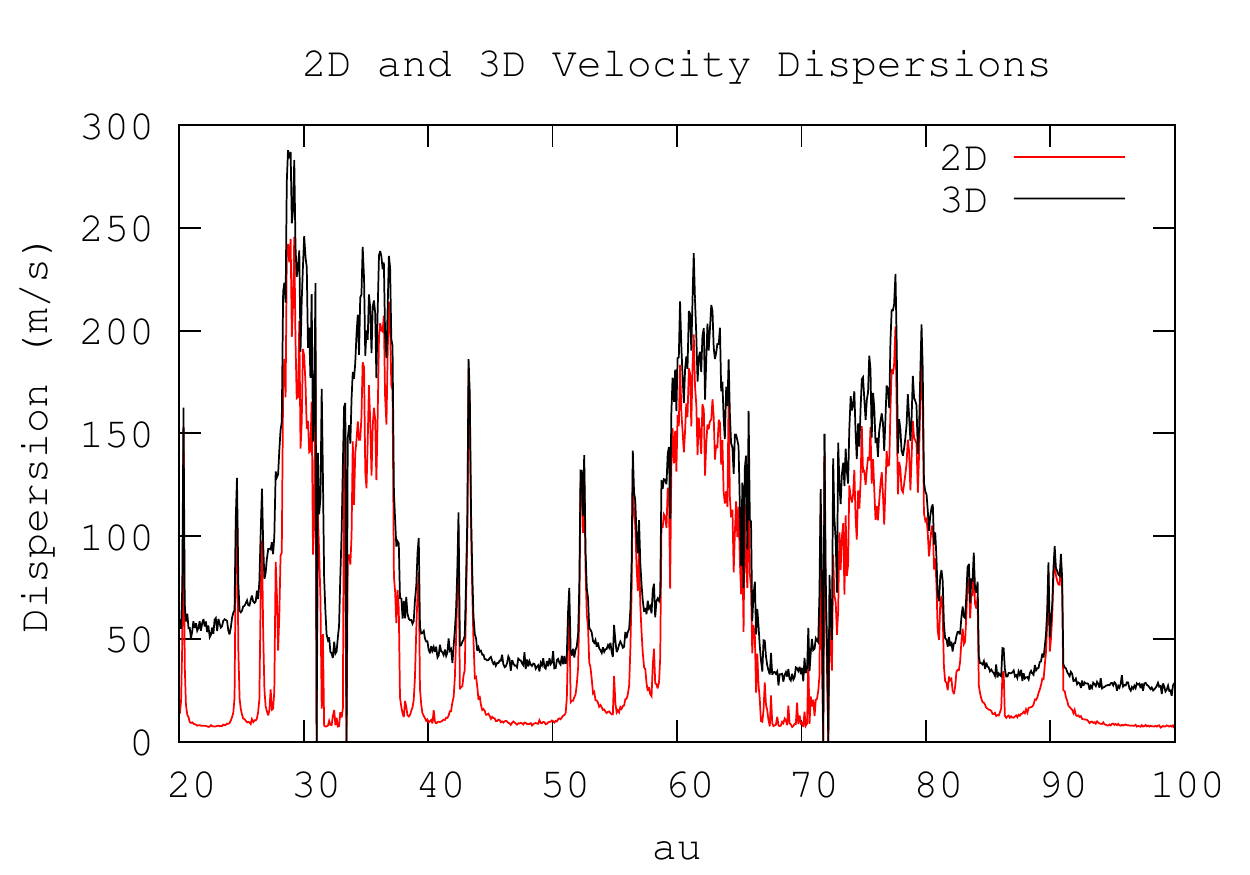}
\caption{\small \ACBc{Collision rates for the planetesimals in the HL Tau, base simulation using the 100,000 yr state. In these plots, all planetesimals are assumed to have an internal density of $1~\rm g~cm^{-3}$ and a radius of 50 km (escape speed of 37 m s$^{-1}$).   {\it Top left.}  The solid curve shows the collision rate as a function of planetesimal semi-major axis using the Safronov-based approach (equation \ref{eqn:simple_collision}).  The red points show the planetesimal collision rates calculated by propagating each particle through the disc for one orbit (equation \ref{eqn:coll3d}).  In this panel, only the radial and azimuthal velocities (and dispersions) are used, which is representative of a disc that is thinner than that explored here (see text).   Collisions rates are suppressed in resonances due to excitation and enhanced in the  cold population due to gravitational focusing.  {\it Top right.}  Similar to the top-left panel, but using the full 3D velocity information.  The high vertical velocity component of the planetesimals prevents gravitational focusing from significantly increasing the collision rates outside of resonances and planet clearing zones.  While the collision rates are lower in the resonances, the relative rate changes are modest compared with those in a thinner disc (top-left panel).  {\it Bottom left.}  The surface density profile for the disc, assuming a total solids mass of $0.001~\rm M_\odot$.  {\it Bottom right.} Velocity dispersion as a function of planetesimal semi-major axis. Both the 2D and 3D dispersions are shown, which together highlight the effects of resonances and the vertical component.     } }\label{fig:collrates}
\end{figure}

The solid lines in the two upper panels of Figure \ref{fig:collrates}  show the results using this method.  We normalize the total mass in solids to be $10^{-3}~\rm M_{\odot}$, which is comparable to the inferred mass of the disc \citep{kwon_etal_2011} with an assumed solid-to-gas mass ratio of 0.01.  
The surface density of solids is shown in the bottom left panel.
All planetesimals are assumed to be identical, with sizes $R=50\rm~km$ and $\rho_p=1\rm~g~cm^{-3}$.  
 The panel on the left assumes only a 2D dispersion (vertical component removed), while the right panel shows the the results for the full 3D dispersion. 
This is done to show how the vertical extent of the disc can impact the collision rate.  
The particles in these simulations were given a random inclination from $0^\circ$ to $1^\circ$, which was motivated initially so that the simulations would be fully 3D.  However, this should have been smaller to be consistent with the mechanism proposed here, as will be discussed momentarily.  
To ensure that using only a 2D dispersion is a reasonable proxy for a thinner disc, a realization of the HL Tau, base  simulation was run with random initial inclinations between $0^\circ$ to $0.1^\circ$ (planets were given inclinations of $0.1^\circ$). The full 3D dispersion for this thinner disc  is consistent with that of the 2D dispersion in regular HL Tau, base simulation,  suggesting that the approximation is justified.    
The 2D and 3D dispersions for the HL Tau, base simulation are shown in the bottom-right panel of Figure \ref{fig:collrates}.

For an additional determination of the collision rates, we use 
\begin{equation}\label{eqn:coll3d}
\dot{N}=\frac{3 \rho_d }{4\rho_p} \frac{R_c^2}{R^3} V_{\rm rel} F_g.
\end{equation}
Evaluating this expression requires an estimate of the spatial volume mass density of planetesimals $\rho_d$.    
To proceed, an azimuthally averaged density structure is calculated by binning particles according to their locations in spherical radius and disc altitude\footnote{We are using disc altitude to refer to the angle measured from the midplane of the disc toward the $z$ axis as seen from the star. This is done to avoid potential confusion with the particle's orbital inclination. }.  
Radial bins are logarithmically spaced between 10 and 200 au with 4096 divisions.  At 45 au, this zoning corresponds to a radial grid width of about 0.033 au.  The disc altitude ranges from $-10^\circ$ to $+10^\circ$, with 512 equally spaced divisions, which yields similar cell sizes in radial width and cell height.  The density $\rho_d$ is then found by assuming that the planetesimals are identical and that the total mass in solids is again $10^{-3}~\rm M_{\odot}$.

The collision rate further requires an estimate of the velocity dispersion in each cell.  As done above, the velocity dispersion is calculated relative to a circular, Keplerian orbit at the particle's cylindrical distance from the star.  With this information in hand, we next take the current orbital elements for each particle and propagate the planetesimal along that fixed orbit for one full period.  This propagation is performed using 1000 equal divisions in true anomaly.  At each position along the orbit, we determine the cell in which the particle is located.  The value for $V_{\rm rel}$ is found by  adding in quadrature the cell's velocity dispersion with the instantaneous velocity of the particle relative to a circular, Keplerian orbit.  We average the values of $V_{\rm rel}$ and $\rho_d$ that correspond to the start and end states for each segment of true anomaly.  The resulting $\dot{N}$ is then used to calculate $p=\dot{N}\Delta t$, where $\Delta t$ is the time step that corresponds to the current change in true anomaly.  The orbit-averaged collision rate is then found by summing $p$ over the entire orbit and dividing by the particle's orbital period. 

The result of this method is also shown in Figure \ref{fig:collrates}.  Each point represents the orbit-averaged collision rate for each particle in the 100,000 yr snapshot.  As with the Safronov-based rate, the left panel shows the results assuming that the vertical components of the velocity can be neglected, while the right panel shows the full 3D result for the given simulation.  Both collision rate calculations, i.e., equations (\ref{eqn:simple_collision}) and (\ref{eqn:coll3d}), are very consistent, with a shift in the rates by a factor of about 3.  Importantly, locations of resonances and planetary chaotic zones show reduced collision rates, although this is modest for the 3D result.  The 2D case shows a large difference between collision rates in and outside of resonances.  This is driven mainly by gravitationally focusing among the dynamically cold planetesimals, which is suppressed at resonances.   The differences between the 2D and 3D cases further show that the planetesimal disc needs to be vertically thin and must have a population of large planetesimals for the proposed mechanism to operate.  Even in the 2D case, if the planetesimals were all about 5 km in radius, then the collision rates would look very similar to the 3D rates shown in Figure \ref{fig:collrates}, just shifted upward due to the increased optical depth of the planetesimal swarm.  

The above calculations establish that a difference in the production of dust is indeed expected near mean motion resonances and chaotic zones, at least for some plausible planetesimal parameters.  
Nonetheless, small grains produced by collisions will drift radially  unless they are confined to their production region.  
We propose that this may be accomplished if the dust grains are reaccreted onto local planetesimals. 
Such accretion is only possible if the reaccretion timescale is less than the migration timescale.  

Let the accretion time of a grain be estimated by 
\begin{equation}\label{eqn:reaccrete_base}
t_{\rm acc}=\left( n \pi R^2 V_{\rm wind} \eta\right)^{-1},
\end{equation}
where $n$ is the number density of planetesimals in the grain's local environment, $R$ is the planetesimal radius for all planetesimals, $\eta$ is an enhancement factor for the geometric cross section, and $V_{\rm wind}$ is the wind speed due to the relative speed between the planetesimals and the gas, which we further assume to be approximately the same as the relative speed between the planetesimals and the grains. 
 For purely gravitational interactions, $\eta = F_g$.  This enhancement factor can also be represented by a pebble accretion-enhanced cross section \citep{hughes_boley2017}, which can allow $\eta>F_g$ for massive enough planetesimals. 
We calculate $n$ by assuming that the midplane planetesimal number density is given by $n=\Sigma_d  /( 2.5 H m_p)$ and that $m_p$ is the typical planetesimal mass.  The value $H$ is the scale height of the vertical planetesimal distribution, which is expected to be smaller than the scale height of the gas disc. 

Using this estimate for $n$ and assuming the typical planetesimal mass is $m_p=\frac{4\pi}{3}\rho_p R^3$, the reaccretion timescale becomes
\begin{equation}\label{eqn:reaccrete_sub}
t_{\rm acc}=\frac{10 \rho_p R H}{3 \Sigma_g V_{\rm wind}\eta}.
\end{equation}
This accretion time must be smaller than the inward drift timescale, which we approximate as $t_{\rm drift} \approx r/V_{\rm drift}$
for disc radial distance $r$ and inward grain drift speed $V_{\rm drift}$.
As discussed in section \ref{sec:method}, the radial drift speed for the mm grains explored here is $V_{\rm drift} = \Delta g t_D$.  
Recall that the residual gravity term $\Delta g$ results from the difference between the gas orbital motion and the local circular, Keplerian orbital speed.  
We set $V_{\rm wind}=\left(1-\xi\right) V_c$, where $\xi$ is a parameterization. 
Using this definition for $V_{\rm wind}$, $\Delta g = 4\pi^2 \left(1-\xi^2\right) r/ T^2 $ for local orbital period $T$ at disc radial distance $r$.  
Altogether, this gives an inward drift speed of
\begin{equation}\label{eqn:radial_drift}
t_{\rm drift} \approx \frac{T^2}{ 4\pi^2 \left(1-\xi^2\right) t_D}.
\end{equation}
We seek a situation such that the ratio $W=\frac{t_{\rm acc}}{t_{\rm drift}}\ll1$.  Using equations (\ref{eqn:radial_drift}) and (\ref{eqn:reaccrete_sub}), 
\begin{equation}\label{eqn:reaccretion}
W=\frac{40\pi^2 \rho_p R H \left(1-\xi^2\right)}{3\Sigma_g V_{\rm rel}\eta}\frac{t_D}{T^2}.
\end{equation}
Letting the Stokes number be $St = 2\pi t_D/T$, equation (\ref{eqn:reaccretion}) can be rewritten as 
\begin{equation}\label{eqn:reaccretionst}
W=\frac{10\rho_p R H \left(1+\xi\right)}{3\Sigma_g \eta }\frac{St}{r}.
\end{equation}
At $r=40$ au, $T\approx 220$ yr, $t_D\sim 0.09$ yr (see section \ref{sec:method}), and $\Sigma_g\sim 0.45\rm~g~cm^{-2}$ (Fig.~\ref{fig:collrates}), the corresponding Stokes number for the mm grains is $St\approx 0.0025$ for these conditions.  
Again setting $V_{\rm rel}=5000~\rm cm~s^{-1}$ as done in section \ref{sec:method}, $\xi\approx0.9906$.  We note that equation (\ref{eqn:reaccretionst}) is not very sensitive to the choice of $\xi$ (and thus $V_{\rm rel}$).
If we let $R=50$ km, $\eta\sim1$, and $H\sim 1000$ km (i.e., $20R$), then $W\sim 0.03$, satisfying the reaccretion/drift condition. 

Taken altogether, the results suggest that the proposed mechanism can operate for plausible planetesimal sizes, provided that the planetesimal disc is geometrically very thin.  It also suggests that, should the proposed mechanism be correct, then the constraints on collision rates along with the small grain reaccretion timescale could be used to infer planetesimal properties.  

\subsection{Extension to the Solar System}

The early debris paradigm presented here could also be reflected in the Solar System's meteoritic record.  
Calcium-Aluminum-rich Inclusions are the oldest known solids in the Solar System \citep{bouvier_wadhwa_2010}, and are presumed to have formed  during or shortly after the onset of the Solar Nebula.  They represent the start of planet building.  
At least some iron meteorite parent bodies formed within 1.5 Myr of the majority of CAIs \citep{schersten_etal_2006}.  
At face value, this suggests that planetesimals large enough to melt and differentiate formed in the Solar Nebula rapidly, potentially on a timescale similar to that of HL Tau.  
Thus, even in the Solar System, there is evidence of very rapid planetesimal \ACBc{and possibly even protoplanetary embryo} formation. 

The proposed model may also be connected with chondrules, which are $\sim0.1$ to 1 mm igneous spherules found in abundance in undifferentiated meteorites \citep{scott_krot_2005}.  
Radiogenic age dating presents evidence that the majority of chondrules formed between 2 and 3 Myr after CAI formation, although some \ACBc{may have} formed contemporaneously with CAIs \citep{villeneuve_etal_2009}.  While chondrule formation is hotly debated, their formation mechanism may be linked to an initial population of planetesimals \citep{hood1998,morris_etal_2012,mann_etal_2016,asphaug_etal_2011, johnson_etal_2015}.
Furthermore, chondrules and planetesimals must have coexisted in the Solar Nebula regardless of their formation mechanism (e.g., certain classes of asteroids are made mostly of chondrules).
This itself has several consequences.  

Any planetesimal that exceeds about 100 km in radius would begin to accrete small solids efficiently through combined gravity and gas drag effects \citep{lambrechts_johansen_2012}.  This pebble accretion would be most efficient for chondrule-sized objects in the regime of the asteroid belt, with some dependence on the disc model \citep{johansen_etal_2015}.  \ACBc{Collisions between chondritic planetesimals would cause mixing and mass migration among the chondrule population and their parent bodies.  This could, in principle, reduce the amount of chondrules that would be lost due to radial drift and could help to explain why chondrites contain chondrules with a range of formation ages \citep[e.g.,][]{villeneuve_etal_2009}. It may also be consistent with recent proposals regarding the types and internal structures of parent bodies \citep{weiss_elkins-tanton2013}.  }

\subsection{Extension to TW Hya }\label{sec:otherdiscs}

\begin{figure}
\includegraphics[width=0.75\textwidth,angle=270]{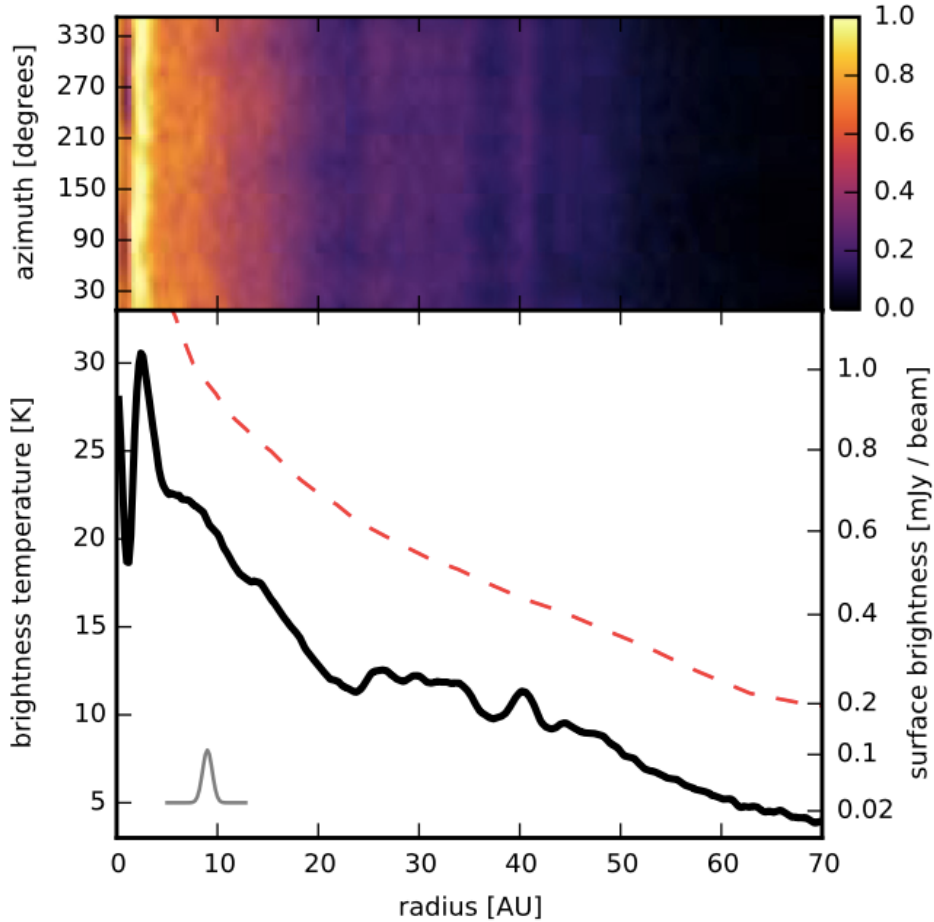}
\caption{Figure 2 from \cite{andrews_etal_2016} showing the radial and azimuthal structure within TW Hya using the high-resolution ALMA 870 $\mu m$ continuum results. The top panel shows the ring and band structure, while the bottom panel shows the azimuthally averaged radial surface brightness profile. The red curve is the midplane temperature profile for a model disc, and the grey profile shows the approximate synthesized beam profile (FWHM$\sim 1$ au).  The large gaps at 22 au, 37 au, and 43 au are clearly seen.  In the averaged profile, the minor bands at about 28 and 31 au can also be seen, although this substructure is harder to see in the top panel.  Following the 43 au band, there is a rapid decline in surface brightness that is not included in this model.}\label{fig:andrewsplot}
\end{figure}

As an additional example, we extend our model and computational methods to TW Hya.  
Observations by \cite{andrews_etal_2016} show that this system harbours wide gaps with some potential band and ring substructure.
The most obvious dark bands  in the Andrews et al.~observations are at $\sim 22$ au, 37 au, and 43 au.  
A bright ring separates the 37 and 43 au bands, again reminiscent of a horseshoe structure.  
However, the disc brightness drops off rapidly after the 43 au band, which is different from that seen in  HL Tau. 
Minor bands in TW Hya appear to be present at 28 au and 31 au, although these are much harder to discern in the current observations.  
There is additional structure at disc radii smaller than 22 au, but as with HL Tau, we ignore this region for computational ease.  
It is straightforward to extend our analysis to these inner regions.  The radial structure of TW Hya as presented by \cite{andrews_etal_2016} is shown in Figure \ref{fig:andrewsplot}.

The 22 au band is broad, and appears to extend from roughly $\sim19$ au to 25 au. 
If a planet is placed at 22 au, then the 3:2 commensurability will be located at roughly 28.8 au, overlapping the minor band near that radial distance.    If an additional planet is located at about 40 au, then the 2:3 commensurability would be at about 31 au, coinciding with the other minor band.  
A planet at 40.5 au would further yield a ring at $\sim40$ au with gaps at about 37 au and 43 au.   

Using the same methods as given in  section \ref{sec:method}, we place a 100,000 test particles on circular orbits between 10 and 60 au. 
A $10~M_{\oplus}$ planet is set at 22 au and a $\sim 19~M_{\oplus}$ planet is placed at 40.52 au, which is at the 5:2 commensurability with the inner planet. 
The strict commensurability may not be necessary.  We place it at this location because the bright ring between the 37 au and 43 au bands is slightly offset from 40 au.  
The masses are chosen to be consistent with the gap widths, assuming a horseshoe structure for the outer gap-ring-gap morphology.  The outer planet is on a circular orbit, and the inner planet has an eccentricity of 0.005. 

\begin{figure}
\includegraphics[width=0.33\textwidth]{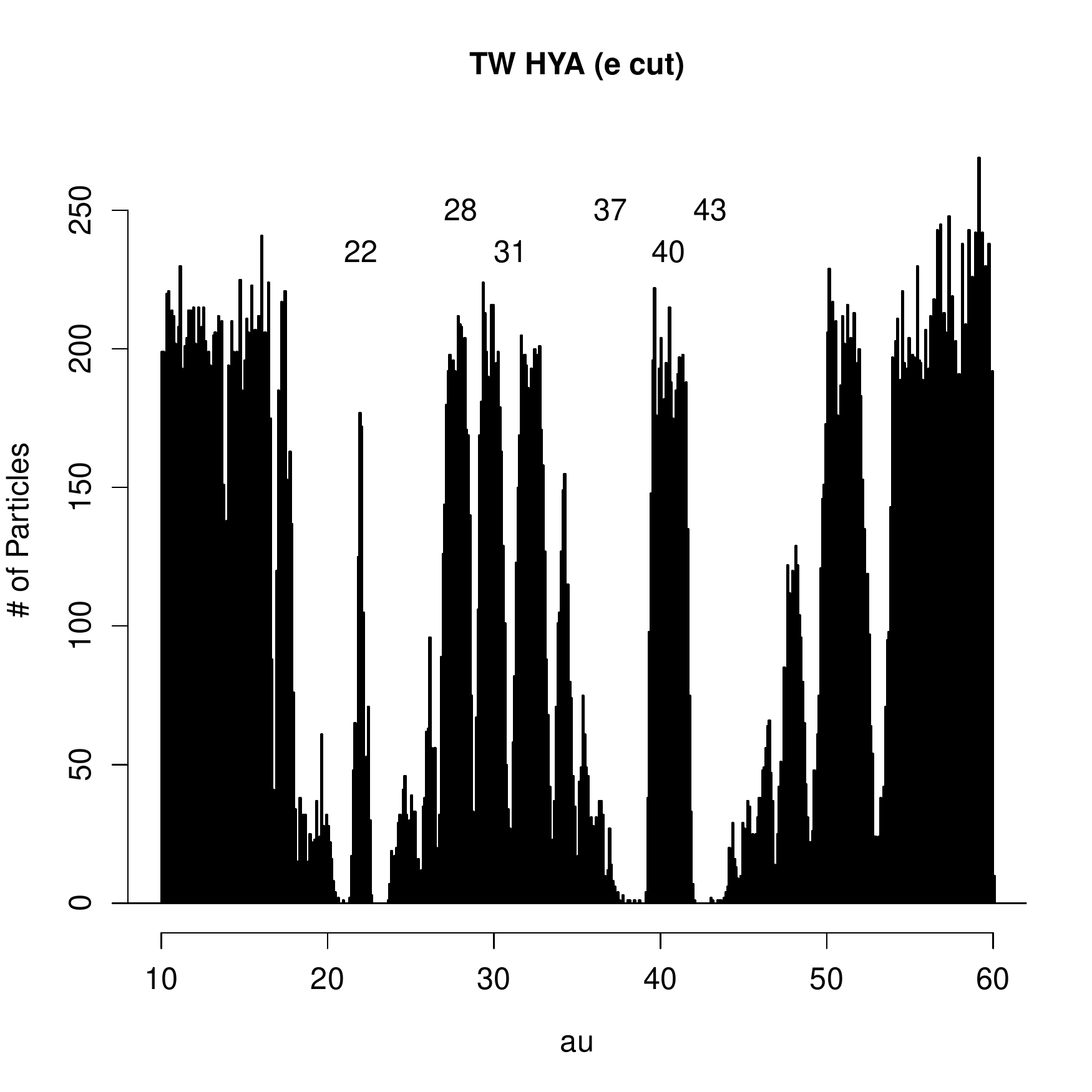}\includegraphics[width=0.33\textwidth]{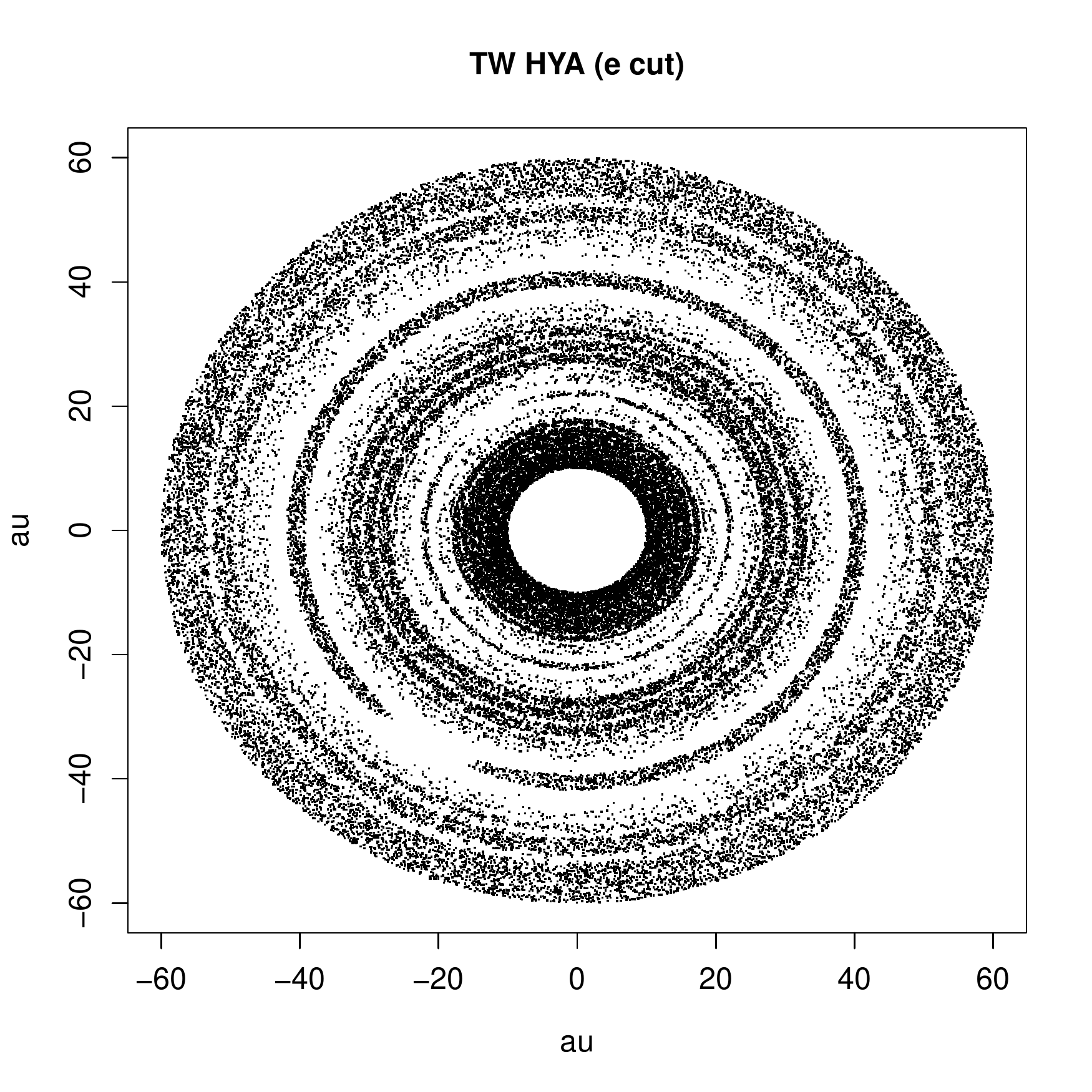}\includegraphics[width=0.33\textwidth]{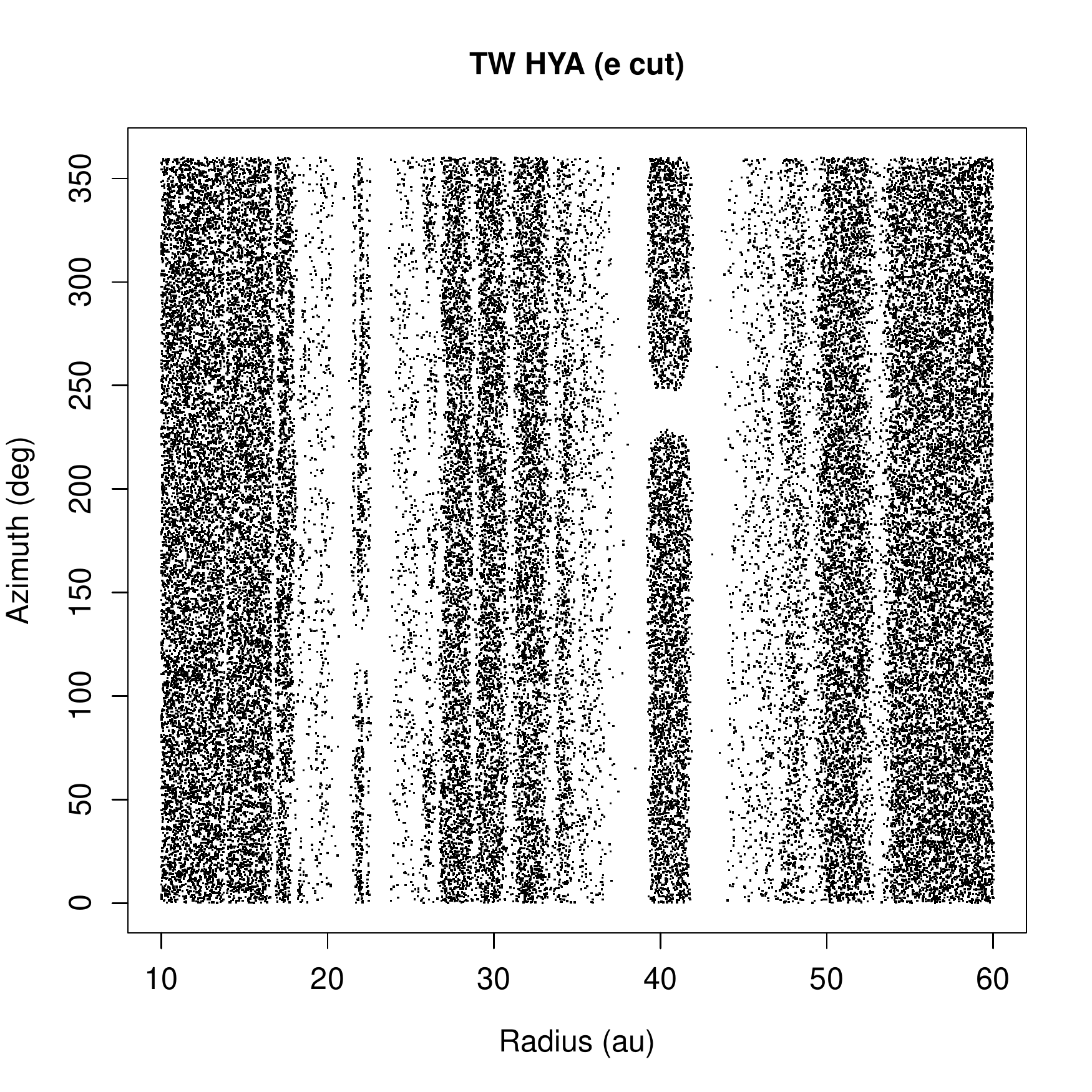}
\caption{Similar to Figure \ref{fig:hltau_hist_ecc}, but representative of TW Hya.  For this simulation, a $\sim 19~M_\oplus$ planet is placed at 40.52 au and  a 10 M$_\oplus$ is placed at 22 au.  The left panel is annotated with the approximate locations of minor and major bands, as identified by \cite{andrews_etal_2016}.  There is a slight shift in the location of the 28 au band (the resonance is at about 28.8 au) and there is an additional potential minor band at 33.4 au due to the 4:3 commensurability with the 40.52 au planet. The middle panel is the face-on image, while the right panel shows the structure in azimuth and radius for better comparison with Figure \ref{fig:andrewsplot}. The overall morphology shares similarities with the actual TW Hya continuum images, although the actual radial surface brightness profile drops off rapidly after 43 au.   Higher sensitivity imaging is necessary to test the dynamical model presented here by clearly identifying the locations of minor bands.  }\label{fig:twhyd_v15_hist_ecc}
\end{figure}

The results are shown in Figure \ref{fig:twhyd_v15_hist_ecc}.  The mild eccentricity of the inner planet produces a wide 22 au gap, while also reducing the amount of low-eccentricity, co-orbital material.  The gap in this particular simulation is slightly too wide to reproduce the TW Hya 22 au band, but this could be altered with the planet's eccentricity.  The 28 au minor band is also produced.  The feature is de-emphasized in the face-on plot compared with the radial binning, but this is also true in the observations \citep[e.g.,][]{andrews_etal_2016}.  The 40.5 au planet produces the co-orbital region as expected, and also produces a minor band at about 31 au. 

The surface brightness profile of the TW Hya mm observations \citep{andrews_etal_2016} shows that the disc brightness never strongly recovers following the 37 au band, which would be inconsistent with the simulation presented here.  However, this may be due to an intrinsic structure of the disc.  Regardless of the reason, additional bands at larger disc radii will be present and coincide with commensurabilities should  the planet hypothesis be correct, unless the entire disc  exterior to 40 au has simply become highly stirred.  
Likewise, the commensurability structure should extend to smaller disc radii, interior to the 22 au planet.  Hints of such structure are present in the existing mm data.  

A potential issue with the proposed paradigm is that the $\sim 40$ au ring in the ALMA continuum image of TW Hya is symmetric, within the fidelity of the observations.  In contrast, the simulations suggest that a distinct, azimuthal horseshoe gap should be present.   We do not have a full explanation for this apparent discrepancy.  Dust within the planet's Hill sphere will partly fill in such gaps, but this alone appears to be insufficient.  For example, the Hill sphere for the proposed 40 au planet is about 1 au, while the horseshoe corotation gap (azimuthal gap)  is about 12-13 au.  This leaves approximately 5 au on either side of the planet that must be accounted for.  This gap could be closed, in part, through gas-drag effects on the mm grains, but this must be tested in simulations that take into account the recycling of mm grains (as proposed here) along with the drag.  Should large azimuthal gaps prove to be required under the proposed mechanism, then high-fidelity mm imaging can test for this signature.


\section{Conclusions}

We have presented a model to explain the presence of gaps in discs.
The points central to the argument are that (1) planetesimals form early and form everywhere in the disc, (2) the mm grains are byproducts of planetesimal collisions and evolution, and (3) the small grains are recycled among the dynamically cold population of the planetesimals, keeping those grains as tracers of that population.  The model will slow the inward migration of small grains, as grains will spend most of their time in larger bodies.  The recycling of material will give rise to pebble-like mass transfer/migration from smaller planetesimals to larger planetesimals and embryos.  Any small solids that become thermally processed while in the nebula (such as chondrules) will have a range of ages, such as seen in meteorites.   \ACBc{When the model is applied to HL Tau, the complex banded structure can be reproduced with three planets (only two are directly modelled), provided the planetesimal disc is geometrically thin.  }

Disc structures are produced through resonances with existing planets.  Planets that are on circular orbits will produce a horseshoe structure with a band-ring-band morphology.  As the planet gains eccentricity, this structure will develop into a single broad band.  
The model predicts that minor bands should be associated with the locations of commensurabilities with proposed planets, provided that the planet is massive enough to cause significant excitation.  This creates a strong test of the model, although there is some flexibility.  For example, eccentric planetary embryos may be able to create deep bands regardless whether they are in or out of a commensurability, although a mean motion resonance may be the most plausible mechanism for forcing an embryo's orbital eccentricity.  

While the model appears to have success in reproducing the morphology of HL Tau and TW Hya, at least at the level of detail presented here, the model must be developed further.  The coevolution of mm grains and planetesimals must ultimately be simulated to establish fidelity between models and the corresponding millimetre continuum imaging.

ACB thanks S.~Andrews, B.~Gladman, P.~Granados, M.~Payne, D.~Tamayo, C.~van Laerhoven, J.~White, and D.~Wilner for discussions that helped to improve this manuscript.  
\ACBc{ACB also thanks the referee, Stuart Weidenschilling, for very helpful comments that improved this manuscript. }
This work was supported by an NSERC Discovery grant, The University of British Columbia, the Canadian Foundation for Innovation, and the BC Knowledge Development Fund.

















\end{document}